\documentclass[%
 reprint,
nofootinbib,
 amsmath,amssymb,
 aps,
]{revtex4-2}

\usepackage{graphicx}
\usepackage{dcolumn}
\usepackage{bm}
\usepackage{xcolor}

\begin{document}

\preprint{APS/123-QED}

\title{Using Cosmic Rays to Improve Weather Forecasting: Meteorological Data Assimilation of Atmospheric Muon Flux Data}

\author{William Luszczak}
\affiliation{%
 Dept. of Astronomy, Ohio State University\\
}%
\affiliation{
Dept. of Physics and Center for Cosmology and Astroparticle Physics, Ohio State University, 191 W. Woodruff Ave, Columbus OH 43201, USA
}%

\author{Man-Yau Chan}
\affiliation{
Department of Geography, 1036 Derby Hall, Ohio State University, Columbus OH 43201, USA
}%

\date{\today}

\begin{abstract}
    Numerical weather prediction requires initial estimates of the atmospheric state. Since the atmospheric density field is intricately woven into the atmosphere's governing equations, advancing atmospheric density estimation will improve numerical weather prediction. However, current meteorological instrumentation cannot directly measure the atmospheric density field over large volumes. Existing techniques rely on sparse point measurements, limiting our ability to accurately estimate the three-dimensional atmospheric density field. One potential solution is to employ measurements of the atmospheric muon flux. Atmospheric muons are particles produced when energetic atomic nuclei (cosmic rays) collide with nuclei in the upper atmosphere, producing a shower of secondary particles (muons) that propagates to the Earth's surface. The surface atmospheric muon flux is known to be proportional to the local atmospheric density field, implying that this technique can be used as a measurement of atmospheric density. This study examines the potential for using atmospheric muon flux measurements to improve atmospheric state estimation via a case study of simulated atmospheric muon observations in the path of tropical cyclone (TC) Freddy. We show that improvement in data assimilation performance can be achieved using data from a relatively small astroparticle detector, well within the capabilities of existing astroparticle technology. Our results also indicate the potential for muon flux measurements to be more effective than surface pressure measurements at state estimation.
\end{abstract}

\maketitle


\section{Introduction}
\subsection{Atmospheric Muons}
Cosmic rays are charged particles (protons, as well as heavier atomic nuclei) accelerated in faraway astrophysical sources. Cosmic rays that reach Earth can interact with nuclei in the atmosphere, producing a shower of secondary particles. Among these particles are muons, produced primarily via pion and kaon decay: 
\begin{equation}
\pi^{\pm} \rightarrow \mu^{\pm} + \overset{\textbf{\fontsize{2pt}{2pt}\selectfont(---)}}{\nu}_{\mu}
\end{equation}
\begin{equation}
K^{\pm} \rightarrow \mu^{\pm} + \overset{\textbf{\fontsize{2pt}{2pt}\selectfont(---)}}{\nu}_{\mu}
\end{equation}

As muons lose energy more quickly when traversing dense matter, the atmospheric muon flux has been observed to be affected by local atmospheric conditions, including pressure, temperature, water vapor content, and precipitation~\cite{LECHMANN2021103842, Jourde_2016, tilav2019seasonal}. This effect, combined with the knowledge that atmospheric muons may travel 10s to 100s of kilometers through the atmosphere before decaying~\cite{Alameddine_2020} suggests that measurements of the muon flux can be used to directly measure the atmospheric density field over large volumes in the vicinity of a muon detector. 

The differential atmospheric muon flux is known to follow roughly a $cos^2(\theta)$ dependence on zenith angle ($\theta$)~\cite{instruments6040078}. While the highest differential muon is in fact from vertical muons ($\theta=0^{\circ}$), geometric phase space effects result in the zenith distribution of observed muons instead peaking near $\theta=30^{\circ}$. In fact, integration of the zenith distribution of observed muons indicates that 50\% of measured muons have zenith angles of $36^{\circ}$ or more. Assuming that atmospheric muon interact at a height of roughly 10 kilometers, this suggests a characteristic distance scale associated with atmospheric muon flux measurements:
\begin{equation}\label{eq:zentodist}
    r_{muon} = 10 \text{ km} \times \text{tan}(36^{\circ}) = 7.27 \text{ km}
\end{equation}
where $r_{muon}$ describes the ``range" associated with the median muon elevation angle. In other words, atmospheric muon flux measurements are roughly 50\% affected by the atmospheric state within 6.49 km, and 50\% affected by the atmospheric state beyond 7.27 km. While this is an oversimplification (we have not taken Earth curvature into account, nor the actual distribution of muon production heights and energies), it does provide an intuition for how muon flux measurements, unlike barometric point measurements, may be able to access atmospheric information beyond their immediate atmospheric vicinity. 

\begin{figure}
    \centering
    \includegraphics[width=\linewidth]{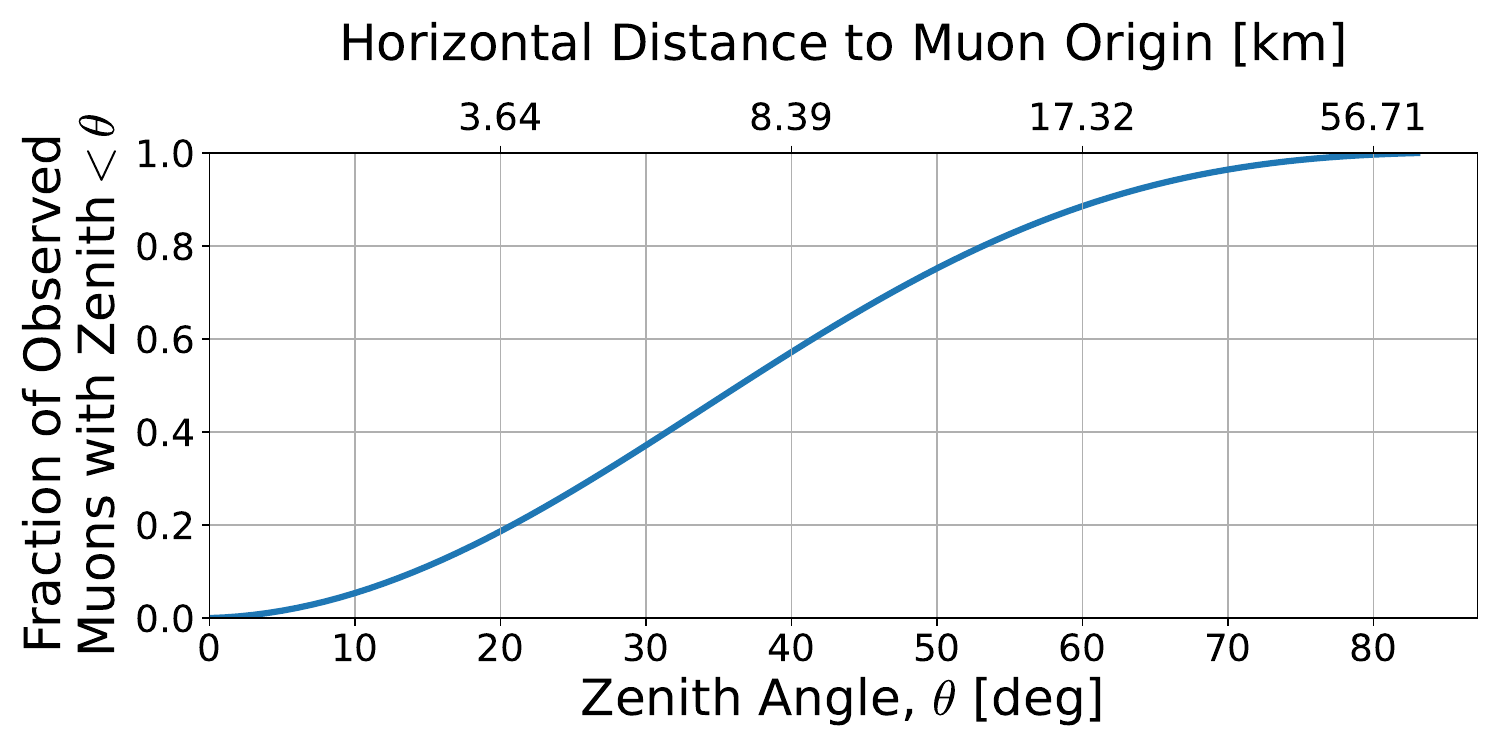}
    \caption{The associated predictive power of an atmospheric muon measurement is associated with the number of muons observed from a particular direction. This plot shows the proportion of muon events that are more vertical than a particular zenith angle, providing an estimation of the relative predictive power if restricting the measurement to only use muons within a particular zenith range. The upper x-axis converts this to a corresponding distance along the Earth's surface, assuming muons are produced at a height of 10 kilometers.}
    \label{fig:muzen}
\end{figure}

Figure~\ref{fig:muzen} shows the relative number of muons expected from zenith angles less than $\theta$. This number is indicative of how much a particular portion of the atmosphere, located a corresponding distance from the detector, affects the overall all-sky integrated muon flux measurement. For the purposes of this plot, we have assumed an average muon production height of 10 kilometers, allowing us to convert zenith angles to horizontal distances as in in equation~\ref{eq:zentodist}. For example, approximately 55\% of the observed muon flux originates from within 8.39 kilometers of the muon detector, suggesting that while the atmospheric state within 8.39 kilometers is the dominant atmospheric factor affecting the muon flux, the portion of the atmosphere between 8.39 kilometers and 56.71 kilometers also has a non-negligible contribution (45\%).

Previous studies have explored variations in the atmospheric muon flux associated with tropical cyclones~\cite{typhoons} and simulated tornadic thunderstorms~\cite{PhysRevD.111.023018}, suggesting that localized atmospheric phenomena do in fact induce a measurable effect on atmospheric muon flux measurements. This in turn suggests the potential to improve weather \textit{prediction} by directly incorporating atmospheric muon flux measurements into numerical weather prediction (NWP) algorithms.

\subsection{Data Assimilation}
NWP systems routinely incorporate atmospheric measurements from many sources to improve the initial conditions used in weather forecasts ~\cite{ECMWF_2024_IFS_DA}. That incorporation is known as data assimilation (DA) and is often presented as a Bayesian inference process ~\cite{Kalnay2003,Fletcher2017ApplicationsGeosciences,Evensen_etal2022_DA_textbook}. Both measurement uncertainties and weather state uncertainties are accounted for in DA. The incorporated measurements range from in-situ measurements (e.g., radiosondes) to remote measurements (e.g., radar backscatter and satellite radiance measurements). As long as a measured quantity has useful statistical associations with atmospheric conditions and the measurement is sufficiently precise, DA has the potential to leverage that measurement to improve NWP initial conditions, and thus forecast accuracy ~\cite{Whitaker2002, Anderson2003AFiltering}. 

\begin{figure}
    \centering
    \includegraphics[width=\linewidth]{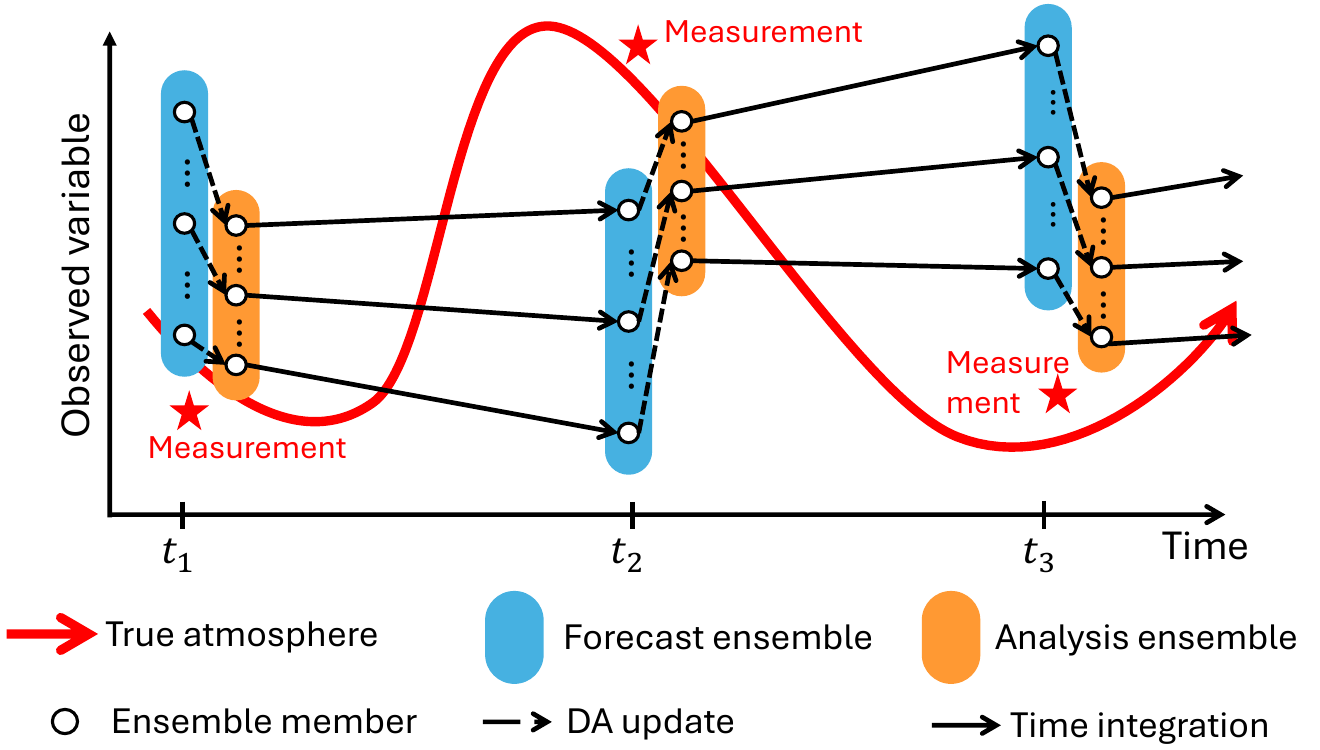}
    \caption{Typical workflow of an EnKF-based DA system. EnKFs begin with an ensemble of forecast model states (blue oval) at the first time ($t_1$) where a measurement are available (red star at $t_1$). Assimilating the measurement updates the ensemble towards the measurement, while reducing the variance of the ensemble (orange oval; ``analysis ensemble"). To assimilate the measurement at the next time (red star at $t_2$), the forecast model is used to time-integrate the analysis ensemble's states at $t_1$ to form the forecast ensemble at $t_2$. After updating the ensemble with that $t_2$ measurement, the $t_3$ measurement is assimilated via the same integrate-then-update process. This integrate-then-update process can be repeated as until the end of the period of interest.}
    \label{fig:ens_DA_illustration}
\end{figure}

This study explores the potential of assimilating muon flux measurements into NWP pipelines via the cutting-edge Data Assimilation Research Testbed (DART) DA system. To be precise, we leverage the ensemble Kalman filter (EnKF) that is available within DART. Fig \ref{fig:ens_DA_illustration} illustrates the workflow of the EnKF. We chose to use the EnKF instead of other DA methods because the EnKF (i) is frequently used in operational NWP, and (ii) can be flexibly adapted to assimilate any observation. For more information, see Appendix A.

Since our study is the first known investigation into assimilating muon flux measurements for NWP, we will use idealized experiments. To be precise, we use Observing System Simulation Experiments (OSSEs) ~\cite{Kalnay2003}, in which a reference weather simulation (henceforth, ``nature run'' or NR) is defined as the ground truth and noisy measurements are generated from the NR. These NR-based measurements are then assimilated into a separate set of weather simulations, and the impact of assimilating muon flux measurements is assessed by examining the difference between that set of simulations and the NR simulation (i.e., the error). We will also determine the muon flux detector exposures needed to achieve reasonable benefits in NWP DA. Finally, we will  compare the impacts of assimilating muon flux measurements against those resulting from assimilating surface pressure measurements.

The rest of this paper is broken into four sections. Section II details the methodology of this study. The results from our experiments are laid out in Section III, and additional discussions are presented in Section IV. This paper concludes with a summary of our findings in Section V. 

Note that the intended audience of this paper is astroparticle physicists. As such, discussions that require expertise on meteorology and/or data assimilation are placed in this paper's appendices instead of the main text.

\section{Methods}

\subsection{Tropical Cyclone (TC) Freddy}

\begin{figure}
    \centering
    \includegraphics[width=\linewidth]{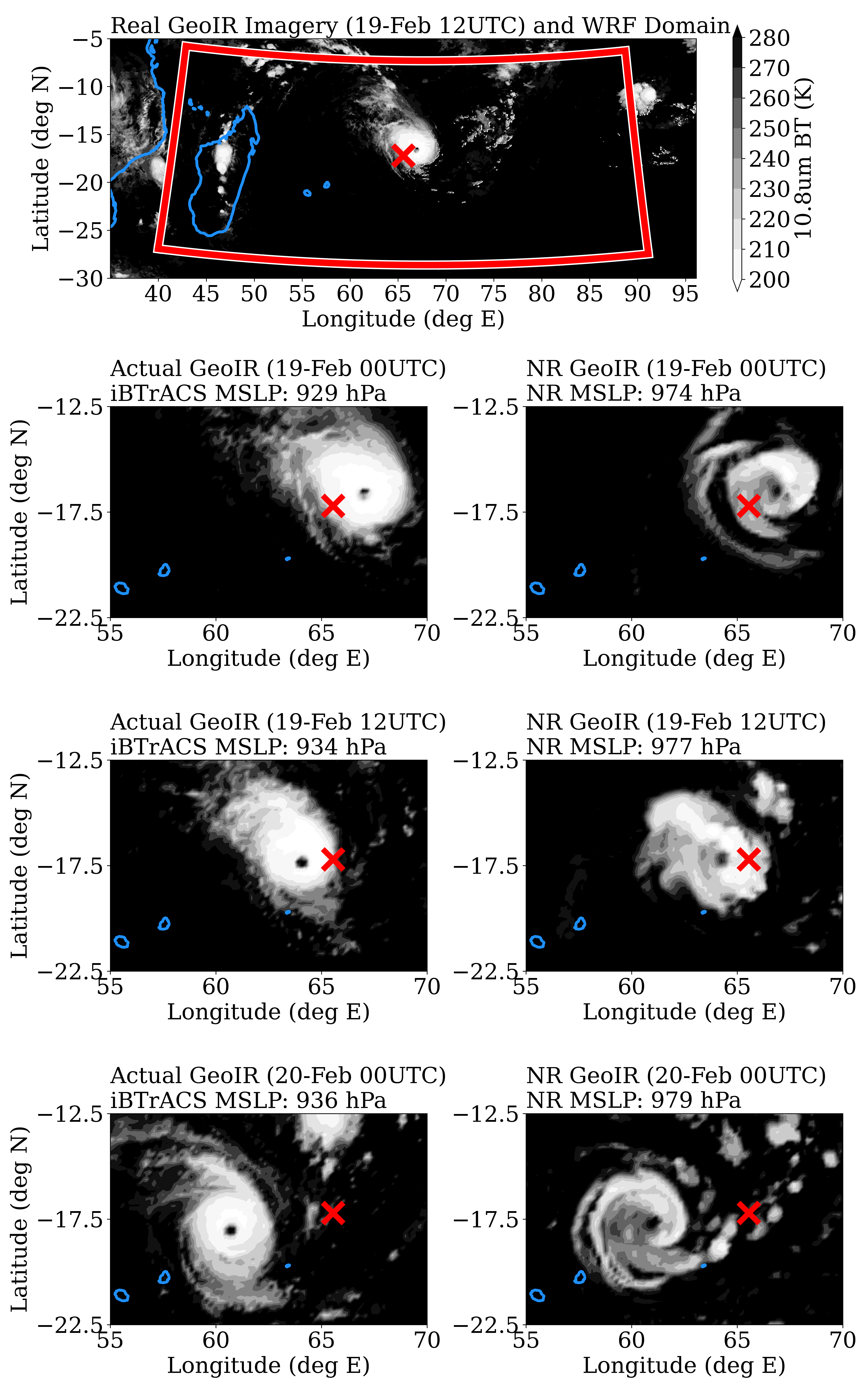}
    \caption{Plots illustrating our setup and nature run. First (top) row: our WRF simulation domain (red) and real 10.8$\mu$m geostationary satellite infrared (GeoIR) imagery on 19 February at 00 UTC. Second row: real GeoIR imagery (left) and NR GeoIR imagery (right) on 19 February at 00 UTC. Third row: real GeoIR imagery (left) and NR GeoIR imagery (right) on 19 February at 12 UTC. Fourth (bottom) row: real GeoIR imagery (left) and NR GeoIR imagery (right) on 20 February at 00 UTC. The iBTrACS and NR TC minimum sea level pressures (MSLP) are indicated in the panels' title in the second, third and fourth rows. Coastlines are indicated in every panel via blue contours. The red cross in every panel indicates the location of our hypothetical muon flux detector.}
    \label{fig:freddy_visuals}
\end{figure}

To explore the efficacy of assimilating atmospheric muon observations, we performed a case study where a hypothetical muon detector is placed in the path of a record-breaking tropical cyclone (TC): Freddy (2023). Freddy is the longest-lasting TC ever recorded worldwide (36 days at or above tropical storm intensity; ~\cite{BAMS2024_TC_Freddy}), caused more than 1,400 fatalities, more than 2,000 injuries, approximately 1.53 billion US Dollars of damages, is the second deadliest TC in the Southwestern Indian Ocean, third deadliest TC in the Southern Hemisphere, and fourth costliest TC in the Southwestern Indian Ocean basin ~\cite{wmo2024_freddy}.

Freddy formed from a tropical low that was first analyzed 5 February 2023 south of Bali ~\cite{BOM_2023_TC_Summary}. Over the course of 17 days, Freddy intensified to the equivalent of a Category 4 hurricane on the Saffir-Simpson scale, occasionally reached the intensity of a Category 5 hurricane, and crossed the Southern Indian Ocean ($\approx$ 7,000 km; \cite{wmo2024_freddy, BOM_2023_TC_Summary}). On 21 February, Freddy made landfall in Madagascar with the intensity of a Category 4 hurricane ~\cite{wmo2024_freddy}. After weakening to a tropical depression while crossing Madagascar, Freddy re-intensified over the Mozambique Channel and made a second landfall in Mozambique on 24 February as a tropical storm ~\cite{wmo2024_freddy}. Freddy then weakened and returned to the Mozambique Channel on 1 March as a tropical depression, before re-intensifying into the equivalent of a Category 1 hurricane, and making a final landfall in Mozambique on 11 March ~\cite{wmo2024_freddy}. Afterwards, Freddy moved inland and dissipated by 14 March ~\cite{wmo2024_freddy}.

\subsection{Setup of NWP Model}

The Weather Research and Forecast (WRF) model version 4.5 is used in this study to produce weather simulations. All of our simulations utilize the same WRF configuration (i.e., our OSSEs are perfect model OSSEs). Our model domain is shown in Figure \ref{fig:freddy_visuals}. For more details and caveats about our WRF setup, see Appendix B. 

To perform OSSEs, a synthetic reality simulation is necessary. This simulation is called the ``nature run'' (NR) and is illustrated by the red curve in Fig. \ref{fig:ens_DA_illustration}. The synthetic observations assimilated in OSSEs are generated from the NR (red stars in Fig. \ref{fig:ens_DA_illustration}) and will be periodically assimilated into our OSSEs.

\subsection{Assessment of the Nature Run}
An assessment of the realism of the NR is a necessary sanity check before performing any DA experiments. Figure \ref{fig:freddy_visuals} compares actual geostationary satellite infrared imagery (left panels) against NR-simulated infrared imagery at three dates within experiments' period. The Radiative Transfer for the Television infrared observation satellite Operational Vertical sounder (RTTOV) model is used to produce the NR-simulated satellite imagery. 

Both the NR-simulated and actual TC Freddy possess an eye. The geographical position of the NR-simulated TC Freddy's eye is within less than 1\textdegree\  off from the actual TC Freddy's eye. Furthermore, both simulation and reality exhibit spiral cloud patterns. As such, the NR replicates reality sufficiently for our OSSEs. 

Note that there are some differences between NR and reality: (i) NR's minimum sea level pressure is roughly 40 hPa higher than reported in the International Best Track Archive for Climate Stewardship (IBTrACS), and (ii) there are fewer high clouds (i.e, low brightness temperature values) in the simulated TC Freddy than in reality. These differences are likely due imperfections in the representation of cloud particle physics \cite{Morrison_etal2020_Review_Cloud_Microphysics}, and our large grid boxes (9-km wide; \cite{Wang2015RegionalResolution}).

\subsection{Atmospheric Muon Simulation}
Muon detectors come in a variety of sizes and intended operation lifetimes, though the relevant quantity for this study is simply the number of observed muon counts observed over a given observation period. We consider a generic, non-tracking muon detector that simply counts the number of muons passing through its interaction plane. Ignoring potential variations in detector response as a function of direction and energy, the number of muon counts observed by a particle detector can be expressed as:
\begin{equation}
\bar{N} = \mathcal{E}_{\text{eff}}\ \times\iint{\Phi(\Omega, E)\  dE\ d\Omega} 
\label{eq:Neq}
\end{equation}
where $\mathcal{E}_{\text{eff}}$ is the ``effective exposure" (units of area $\times$ time), characterizing the physical size, average response, and lifetime of the particle detector, and $\Phi(\Omega, E)$ is used to denote the atmospheric muon flux at energy $E$ from direction $\Omega = (\theta, \phi)$. $\theta$ and $\phi$ are angles describing the elevation and azimuthal direction of a portion of the sky in the coordinate system centered on the particle detector. For the purposes of this study, $\phi=0^{\circ}$ points to the North, and increases in the clockwise direction. 

Atmospheric muon fluxes are numerically calculated using MCEq\footnote{\url{https://github.com/mceq-project/MCEq/tree/master/MCEq}}, a numerical tool for solving cascade equations that model the evolution of particle densities as they traverse gaseous media. For the purposes of this study, MCEq is modified to incorporate atmospheric density profiles loaded directly from external atmospheric simulation, allowing for the calculation of the atmospheric muon flux under the nature run atmosphere, as well as all 50 ensemble member atmospheres. The density field information imported into MCEq is obtained directly from the WRF simulation, and as such accounts for the dry-air density as well as water vapor. Clouds and suspended precipitation are not included in the density field, however these only account for a miniscule fraction of the total density ($<$1 g/kg), and previous simulation studies of the effect of hydrometeors on the atmospheric muon flux indicate that the effects of precipitation are subdominant to dry-air density variations~\cite{PhysRevD.111.023018}. The primary cosmic ray flux is simulated according to the model described in \cite{h3acitation} and primary cosmic ray interactions are then simulated using the SIBYLL 2.3c interaction model~\cite{sibyll}. Muons resulting from these interactions are propagated through the relevant atmospheric density field to a hypothetical muon detector placed at sea level at (longitude, latitude) = $(\lambda, \psi) = (65.55^\circ, -17.21^\circ)$, resulting in a direction and energy-dependent muon flux in the coordinate system of the muon detector, $\Phi(\Omega, E)$.

For a simulated muon detector with a given effective exposure, an observed number of muon counts ($N_{obs}$) is generated by drawing a random value from a Poisson distribution with mean $\bar{N}$:
\begin{equation}\label{eq:muon_poisson}
    \tilde{P}(N_{obs}) = \frac{\bar{N}^{N_{obs}} e^{\bar{N}}}{N_{obs}!}
\end{equation}
where $\bar{N}$ is calculated from the directional muon flux using equation~\ref{eq:Neq}. As $N_{obs}$ is correlated with local atmospheric properties, we can treat $N_{obs}$ as a meteorological measurement (with measurement error $\sqrt{N_{obs}}$). This information can then be passed to a data assimilation framework to assimilate a muon flux observation for a particular atmospheric ensemble member at a given time.

\subsection{DA System}
The National Center for Atmospheric Research's Data Assimilation Research Testbed ~\cite{Anderson2009TheFacility} (DART\footnote{\url{https://dart.ucar.edu/}}; Manhattan release) is used to assimilate measurements in this study. Specifically, we use DART's two-step Ensemble Adjustment Kalman Filter (EAKF; \cite{ Anderson2003AFiltering, Anderson2001AnAssimilation}). All DA experiments in this study employ 50-member WRF ensembles. Each member is (conceptually) a Monte Carlo sample drawn from an underlying prior distribution of weather states (white circles in Fig \ref{fig:ens_DA_illustration}). For more details of our DA setup, see Appendix C. 



\subsection{Setup of OSSEs}
A series of six OSSEs are conducted to investigate the potential impacts of assimilating muon flux measurements into NWP. The first OSSE (henceforth, "NoDA") is a control experiment that assimilates no observations. To be clear, the NoDA OSSE consists of 50 freely-evolving WRF ensemble members and no DA whatsoever is performed. The next four OSSEs assimilate muon flux observations every hour at the center of the simulation domain (17.21\textdegree S, 65.55\textdegree E), albeit with differing detector exposure: 10\textsuperscript{3} m\textsuperscript{2}s, 10\textsuperscript{4} m\textsuperscript{2}s, 10\textsuperscript{5} m\textsuperscript{2}s, and 10\textsuperscript{6} m\textsuperscript{2}s. The final OSSE assimilates a surface pressure (PSFC) measurement every hour at the center of the domain (17.21\textdegree S, 65.55\textdegree E). 

All assimilated measurements in this study are synthetically generated from the NR. To be precise, suppose the function $h$ denotes obtaining an error-free measurement from an NR atmospheric state $\boldsymbol{x_{*}}$. Any measurement $y_o$ assimilated in this study account for measurement uncertainties via
\begin{equation}
    y_o = h\left(\boldsymbol{x_{*}}\right) + \epsilon_o
\end{equation}
where $\epsilon_o$ is a random sample drawn from the measurement error distribution. Every muon flux measurement's $\epsilon_o$ is drawn from the Poisson distribution defined in Eq. \eqref{eq:muon_poisson}, and every PSFC measurement's $\epsilon_o$ is drawn from a normal distribution with a mean of zero and a standard deviation of 100 Pa.

\section{Results}

\subsection{Definition of DA performance metrics}
We use normalized root-mean-squared error (nRMSEs) to assess the potential impacts of assimilating muon flux and surface pressure observations for NWP. To define nRMSEs, it is necessary to first define root-mean-square-error (RMSE). The RMSE for a particular weather model quantity $x$ is defined as:
\begin{equation}\label{eq_rmse}
    \text{RMSE} := 
    \sqrt{
        \dfrac{1}{N_{loc}}
        \sum^{N_{loc}}_{\ell=1}
        \left( 
            \overline{x_\ell} - x_{*,\ell}
        \right)^2
    }
\end{equation}
$N_{loc}$ is the number of latitude-longitude locations the average is performed over, $x_{*,\ell}$ denotes the true value at grid box $\ell$, and $\overline{x_\ell}$ denotes the ensemble mean value at grid box $\ell$. $\overline{x_\ell}$ is defined to be:
\begin{equation}
    \overline{x_\ell} 
    :=
    \dfrac{1}{N_{ens}}
    \sum^{N_{ens}}_{n=1}
    x_{n,\ell}
\end{equation}
where $x_{n,\ell}$ denotes the value of the $n$-th ensemble member at grid box $\ell$.

The normalized RMSE (nRMSE) is defined as
\begin{equation}
    \text{nRMSE}:= 
    \dfrac{
        \text{OSSE RMSE}
    }{
        \text{NoDA RMSE}
    }
\end{equation}
nRMSE values below 1 indicate that the OSSE outperforms NoDA and \textit{vice versa} for nRMSE values exceeding 1. 

The nRMSEs of five meteorological fields are examined in our assessment: surface pressure (PSFC), eastward wind velocity component (U), northward wind velocity component (V), potential temperature (T), and water vapor mixing ratio (QVAPOR). Note that because U, V, T and QVAPOR are three-dimensional scalar fields, their nRMSEs are functions of both time and altitude. As such, we summarize those four fields' nRMSEs using averages over pseudo-pressures. Pseudo-pressure $p$ is defined via
\begin{equation}
    p 
    := 
    \left(P_0 - P_{top}\right) \,\eta + P_{top}
\end{equation}
where $P_0$ is a constant set to the standard surface pressure value of 10\textsuperscript{5} Pa, $P_{top}$ is the constant model top pressure of 2000 Pa, and $\eta$ denotes the terrain-following vertical coordinate used by WRF. The time-varying $p$-averaged nRMSE is thus defined as
\begin{equation}
    \left\langle nRMSE \right\rangle_p \left(t\right)
    :=
    \dfrac{
        \int^{P_0}_{P_{top}} nRMSE\left(p,t\right) \, dp
    }{P_0-P_{top}}
\end{equation}
where $t$ denotes time. The time-varying $p$-averaged nRMSEs are displayed in Figure \ref{fig:rmse_uvtq_plot}.

\subsection{The Impacts of Assimilating Muon Flux Measurements}

\begin{figure}
    \includegraphics[width=0.45\textwidth]{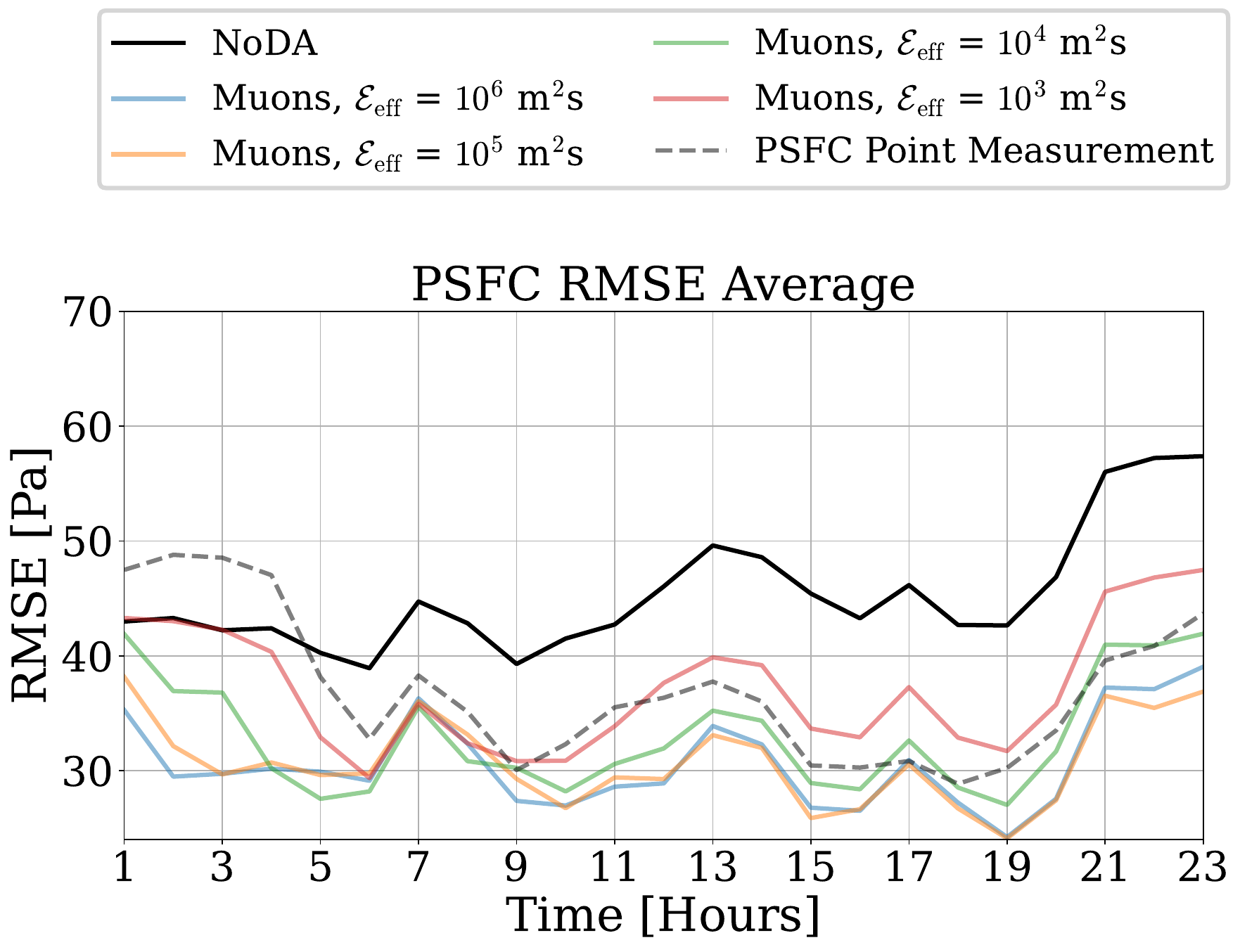}
    \includegraphics[width=0.45\textwidth]{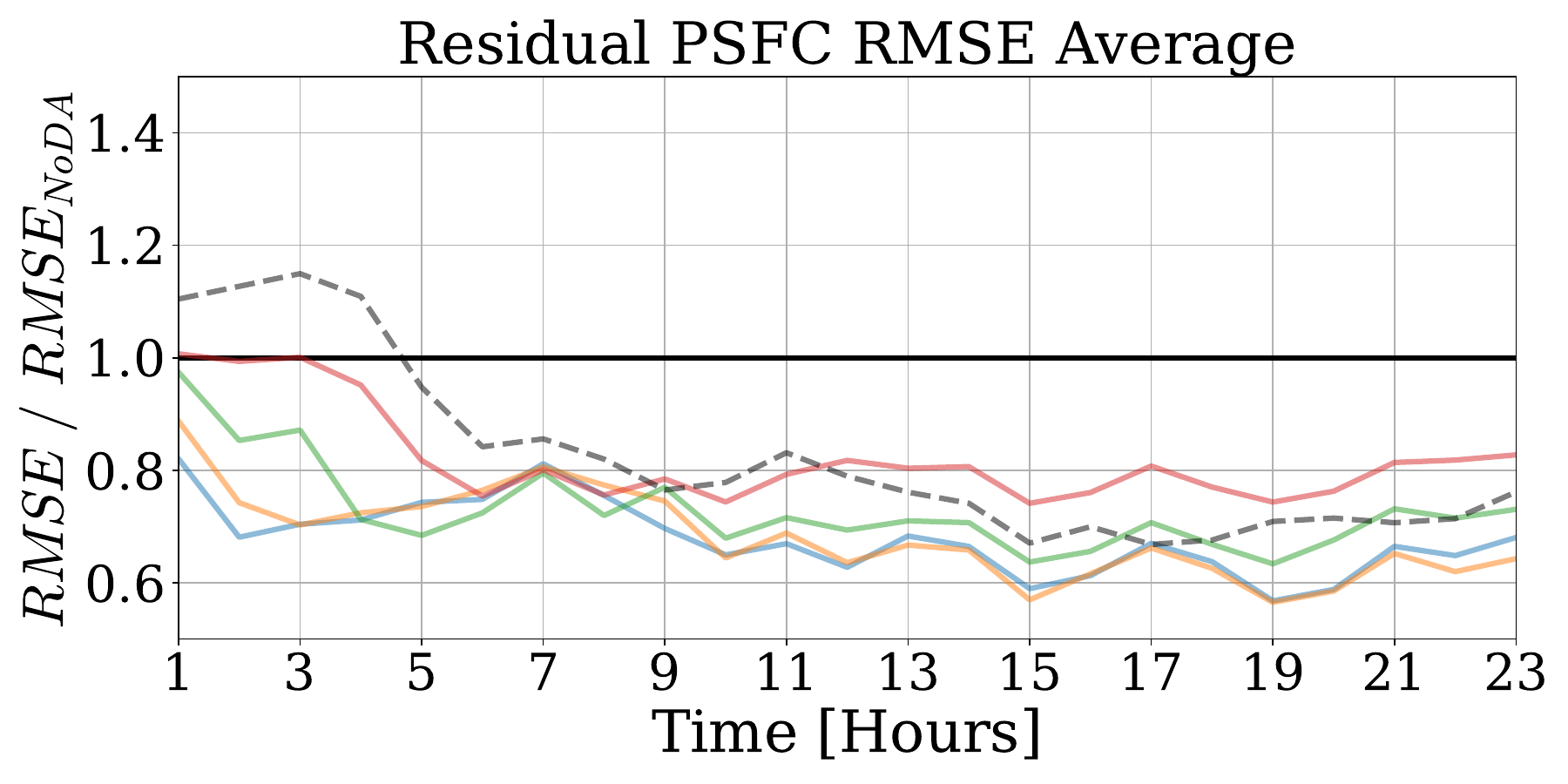}
    \caption{Top: Surface pressure (PSFC) RMSE values, calculated at each time step of data assimilation, averaged over the full domain (longitude ranging from 39.17$^{\circ}$ to 91.12$^{\circ}$, latitude ranging from -28.63$^{\circ}$ to -5.78$^{\circ}$). Colored lines represent simulated muon detectors with various effective exposures, ranging from $10^3$ m$^2$s to $10^6$ m$^2$s. For comparison, the free evolution of the atmospheric model (``NoDA") is plotted as a black line. Assimilation of a single surface pressure measurement at the location of the muon detector is shown as the dashed black line. Bottom: The residuals of the hourly RMSE values relative to the ``NoDA" case of free model evolution.}
    \label{fig:rmse_psfc_plot}
\end{figure}

\begin{figure}
    \includegraphics[width=0.5\textwidth]{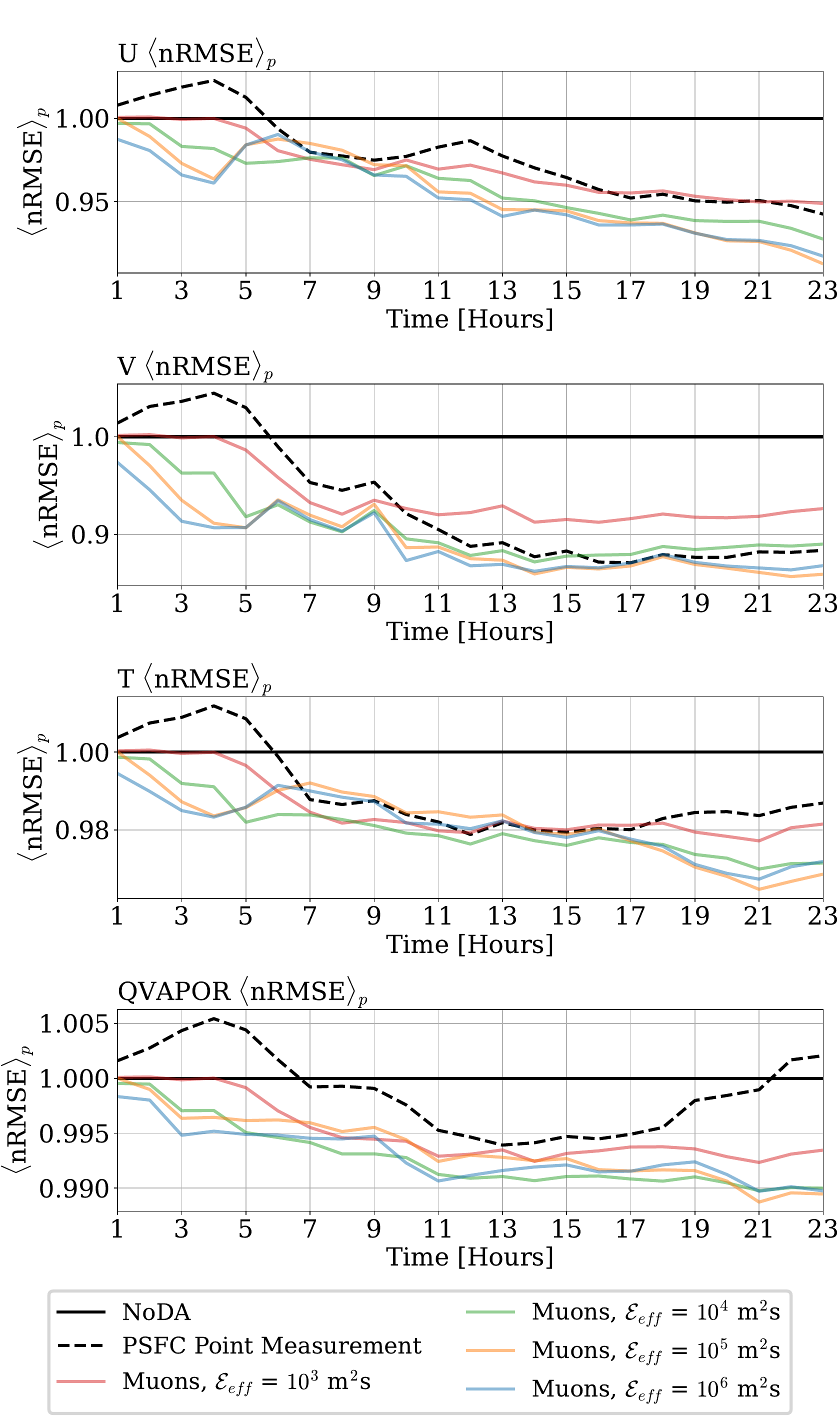}

    \caption{Similar to Fig 3, except that the PSFC RMSEs are replaced with the domain-wide $\left\langle\text{nRMSE} \right\rangle_p$ of U (top row), V (second row), T (third row), and QVAPOR (bottom row).}
    \label{fig:rmse_uvtq_plot}
\end{figure}

All four OSSEs that assimilated muon flux measurements produced RMSEs that are smaller than NoDA's (Figures \ref{fig:rmse_psfc_plot} and \ref{fig:rmse_uvtq_plot}). Compared to NoDA, those OSSEs' PSFC RMSEs are between 20--40\% smaller, U RMSEs are up to 15\% smaller, V RMSEs are up to 23\% smaller, T RMSEs are up to 4\% smaller, and QVAPOR RMSEs are up to 3\% smaller. The impacts of muon flux DA on time-averaged nRMSEs as a function of $p$ are available in Appendix D. The corrective impacts of muon flux DA on T and QVAPOR are likely explained by the ideal gas law -- T and QVAPOR influence air density, which is then sensed by muon flux measurements. This physical relationship is likely reflected in the ensemble statistics that the EAKF DA algorithm leveraged to convert the muon measurement into model state adjustments.

To understand the corrective impacts of assimilating muon flux measurements on U, V and PSFC  we examined the adjustments to the ensemble mean PSFC, U and V resulting from assimilating the muon flux measurement at the 6 hour mark in the $10^6$ m$^2$s OSSE. These adjustments are called ``analysis increments'' in the DA literature. 

\begin{figure}
    \centering
    \includegraphics[width=0.4\textwidth]{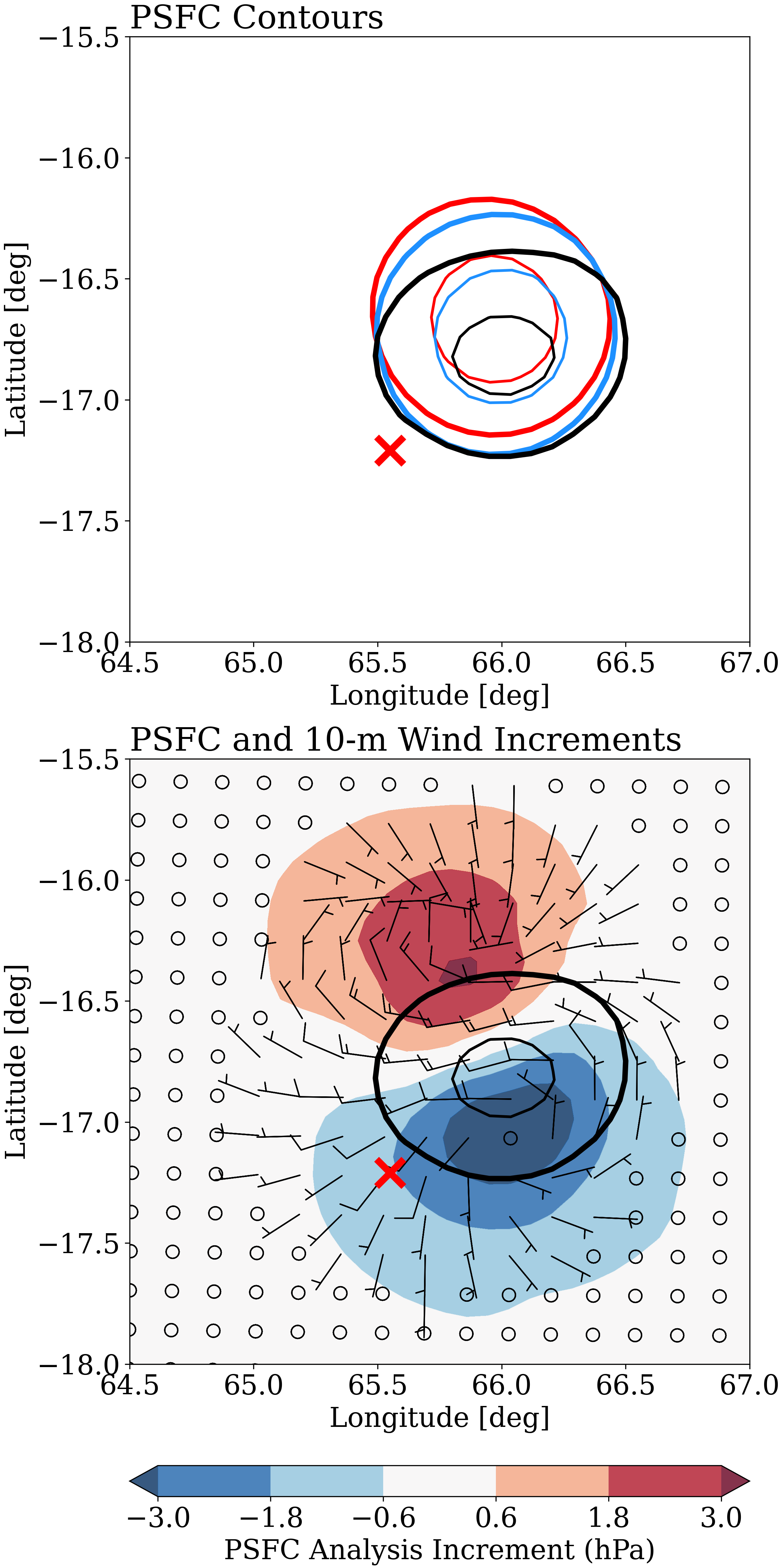}
    \caption{Top: PSFC contours corresponding to 980 hPa (thin) and 990 hPa (thick) obtained from the nature run and $10^6$ m$^2$s OSSE at the 6 hour mark (i.e., 19 Feb at 06 UTC). The red contours correspond to the forecast ensemble mean PSFC, the blue contours correspond to the analysis ensemble mean PSFC, and the black contours correspond to the nature run's PSFC. Bottom: PSFC analysis increments (shadings) and 10-meter horizontal wind velocity analysis  increments (wind barbs in units of knots) from the $10^6$ m$^2$s OSSE at the 6 hour mark. The black contours are the same as the top panel. Red crosses indicate the muon flux observation site.}
    \label{fig:ana_inc}
\end{figure}

TCs are associated with low PSFC, and are thus identifiable from PSFC contours (a.k.a., surface isobars). We use this association to parse the positions of the simulated TC Freddy from PSFC. The top panel in Figure \ref{fig:ana_inc} indicates that the low PSFC values in the forecast ensemble mean are displaced north of the nature run's low PSFC values. In other words, the TC in the forecast ensemble is displaced north of the nature run's TC. The assimilation of a muon flux measurement with $10^6$ m$^2$s exposure causes the ensemble mean's low PSFC contours to shift towards the nature run's (top panel of Figure \ref{fig:ana_inc}), implying that that assimilation corrected to the general position of TC Freddy in the ensemble.

The correction to the placement of the ensemble mean's low PSFC contours is a result of the dipolar surface pressure analysis increment imposed by muon flux EnsDA (bottom panel of Figure \ref{fig:ana_inc}). To be precise, EnsDA imposed a positive PSFC analysis increment on the northern side of the forecast ensemble mean low PSFC contours and a negative PSFC PSFC increment on the southern side of the forecast ensemble mean low PSFC contours. 

Figure \ref{fig:ana_inc} (bottom panel) also illustrates how muon flux DA improved the performance of U and V. Analysis increments to U and V at an altitude of 10 m are consistent with the gradient wind under friction: the diverging anticyclonic increments are broadly collocated with the positive PSFC increment and the converging cyclonic increments are broadly collocated with the negative PSFC increment. In other words, muon flux EnsDA improves the U and V fields by leveraging ensemble statistics that reflect geophysical airflows.  

Note: the asymmetry in the PSFC analysis increment dipole (the southern half's magnitude is stronger than the northern half's) is likely a result of the covariance localization used to ameliorate sampling errors (see Appendix C).

In summary, our OSSEs demonstrate that muon flux measurements have the potential to improve NWP DA.

The more interesting question is: are muon flux measurements more useful for NWP DA than PSFC measurements? This question is particularly interesting because PSFC is related to atmospheric density via hydrostatic balance. Both muon flux measurements and PSFC measurements are thus similar in the sense that both measurements are related to atmospheric density. 

Out of the four OSSEs that assimilated muon flux measurements, three of them (exposures exceeding $10^3$ m$^2$s) have nRMSEs that are smaller than the OSSE that assimilated PSFC measurements. Compared to the PSFC-assimilating OSSE, these three muon-assimilating OSSEs have PSFC nRMSEs that are up to 0.1 smaller. Since the latter OSSE's PSFC nRMSEs are roughly 0.3 smaller than NoDA, the assimilation of muon flux measurements are up to 30\% more effective for improving PSFC than the assimilation of PSFC measurements. Applying similar logic to Figure \ref{fig:rmse_uvtq_plot} reveals that assimilating muon flux measurements, compared to assimilating PSFC measurements, are up to 70\% more effective for improving U, up to 20\% more effective for improving V, and up to 100\% more effective for improving T. 

Note that the OSSE that assimilated PSFC measurements performed worse than the NoDA OSSE for the first few hours (Figures \ref{fig:rmse_psfc_plot} and \ref{fig:rmse_uvtq_plot}). A likely explanation for this behavior is that the stochastic nature of measurement noise randomly caused the PSFC measurements to be less accurate than the ensemble mean. Future work can investigate countermeasures against such situations (e.g., quality control).

The performance of the PSFC-assimilating OSSE and the muon-assimilating OSSE diverge for QVAPOR. All four muon-assimilating OSSEs have $\left\langle\text{nRMSE}\right\rangle_p$ less than unity, whereas the PSFC-assimilating OSSE's QVAPOR $\left\langle\text{nRMSE}\right\rangle_p$ hover around unity. In other words, the assimilation of muon flux measurements improves the humidity field whereas the assimilation of PSFC results in neutral impacts. 

The fact that the assimilating muon flux measurements produces better PSFC fields than assimilating PSFC measurements is interesting. To better understand that fact, we examine the squared error between the ensemble average and the true value at a grid box $\ell$ within the domain. The squared error
\begin{equation}
    \delta_\ell^2 := (\overline{x_\ell} - x_{*,\ell})^2
\end{equation}
characterizes the ensemble mean's accuracy at a particular location (i.e., grid box $\ell$). Because $\delta_\ell^2$ is literally the summand in RMSE (Eq. \eqref{eq_rmse}), $\delta_\ell^2$ can be investigated to determine which parts of the domain most contribute to improved or worsened RMSE values. 

\begin{figure}
    \includegraphics[width=0.45\textwidth]{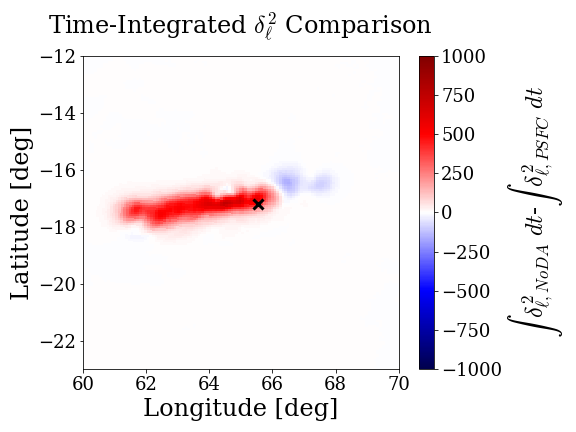}
    \includegraphics[width=0.45\textwidth]{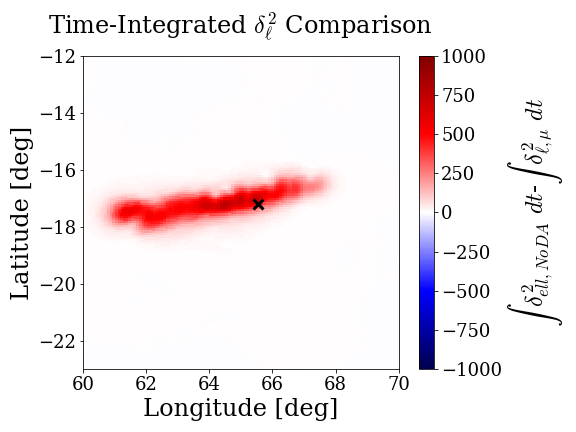}
    \includegraphics[width=0.45\textwidth]{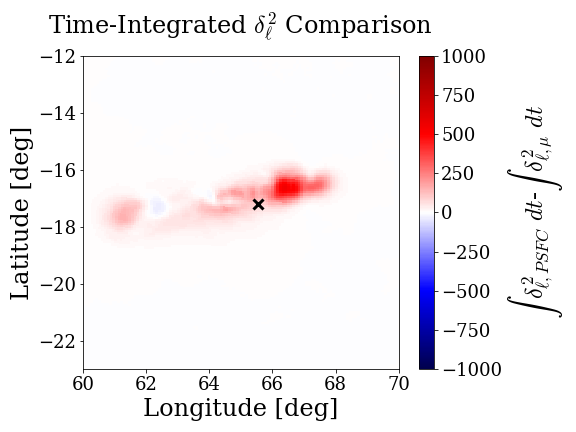}
    \caption{Top: The difference in surface pressure RMSE between free evolution (``NoDA") and assimilation of a single surface pressure point measurement (``PSFC"), integrated over 24 hours.  Middle: The same plot, but assimilating muon flux instead of using a surface pressure point measurement. Bottom: The difference between assimilation using surface pressure point measurements and using muon flux information. Red regions in the bottom two plots are where muon flux assimilation has improved surface pressure prediction ability (the RMSE when assimilating muon flux data is \textit{smaller} than the free evolution/surface pressure assimilation case), while blue regions are where muon flux assimilation produces worse predictions. The simulated observation position is denoted by a red ``X".}
    \label{fig:rmse_tint_plot}
\end{figure}

Figure \ref{fig:rmse_tint_plot} compares the time-integrated impacts on $\delta_\ell^2$ resulting from (i) assimilating PSFC measurements and (ii) assimilating muon flux measurements with an exposure of $10^5$ m$^2$s. While both experiments reduced PSFC $\delta_\ell^2$ in a time-integrated geographically-averaged sense, the former OSSE degraded the accuracy of the PSFC field to the east of the observation site whereas the latter OSSE showed no clear regions where the PSFC field is degraded. Furthermore, both OSSEs have similar performances to the west of the observation site. As such, the differences in PSFC RMSEs are largely due to performance differences to the east of the observation site.

That unique region is likely because muon flux measurements are affected by atmospheric conditions over a large volume surrounding the observation location, not just the column density directly above the detector. Muon flux measurements can consequently obtain information about atmospheric density perturbations before weather systems propagate directly over the measurement location, while surface pressure point measurements only probe the atmospheric conditions at the location of the measurement device.

In summary, our OSSEs indicate that (i) muon flux measurements have the potential to improve NWP and (ii) that potential is greater than that of PSFC measurements when the muon flux measurement exposure exceeds $10^3$ m$^2$s. We note that this mirrors the physical source of both these observations: surface pressure is related to the integrated vertical profile of atmospheric density via hydrostatic balance, while the atmospheric muon flux is also directly affected by the atmospheric density field as described in section II.B. While similar, the differences between these measurements, including the non-locality of muon flux observations, contribute to different performance when assimilating the observations considered here. For example, for all exposures considered in this study, the assimilation of muon flux measurements have a stronger potential to increase the accuracy of humidity fields than PSFC measurements.

\section{Additional Discussion}
\subsection{Muon Detector Size Requirements}

\begin{figure}
    \includegraphics[width=0.4\textwidth]{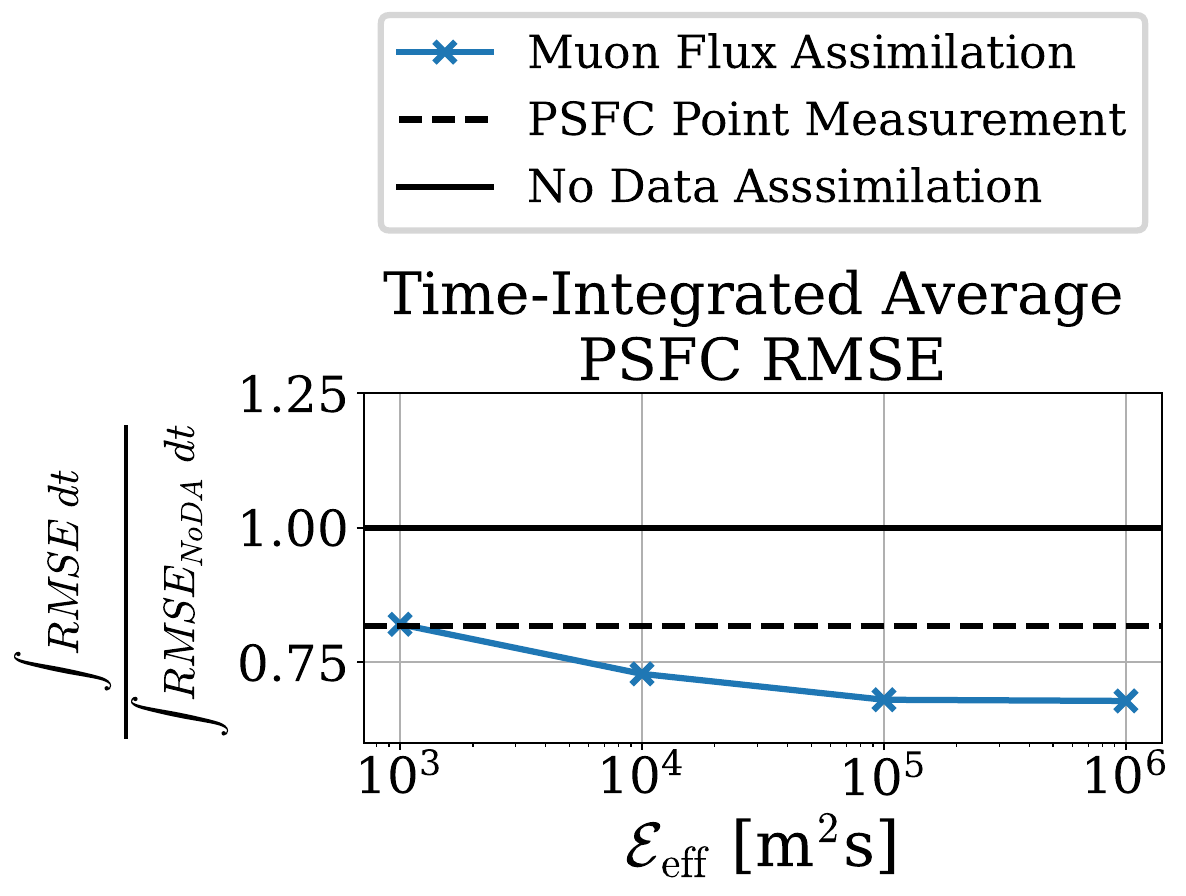}
    \includegraphics[width=0.4\textwidth]{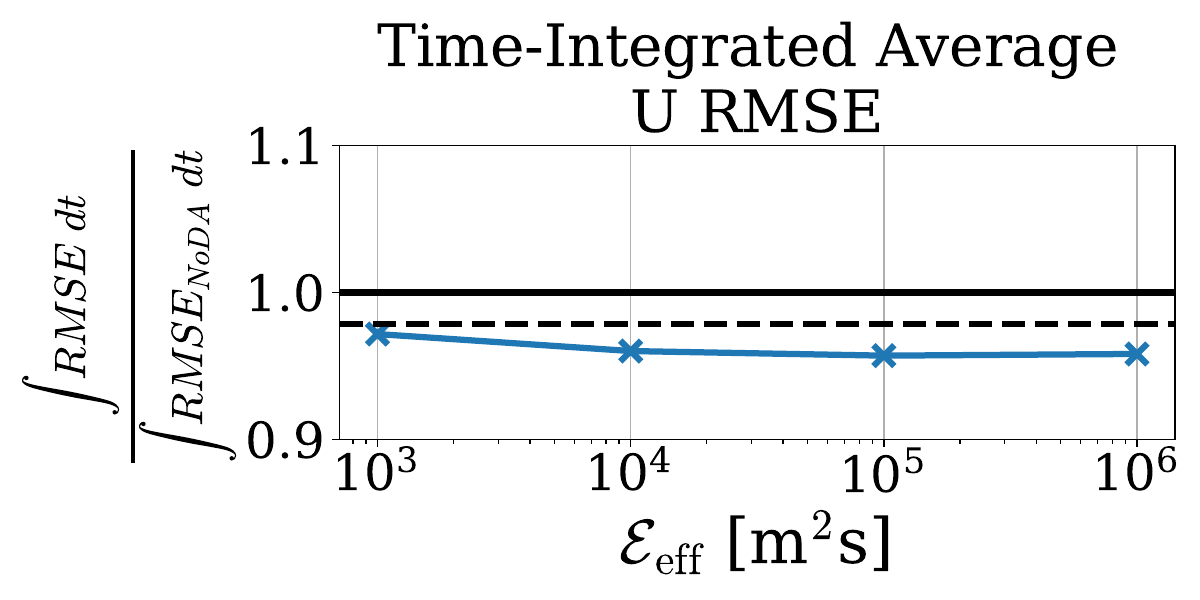}
    \includegraphics[width=0.4\textwidth]{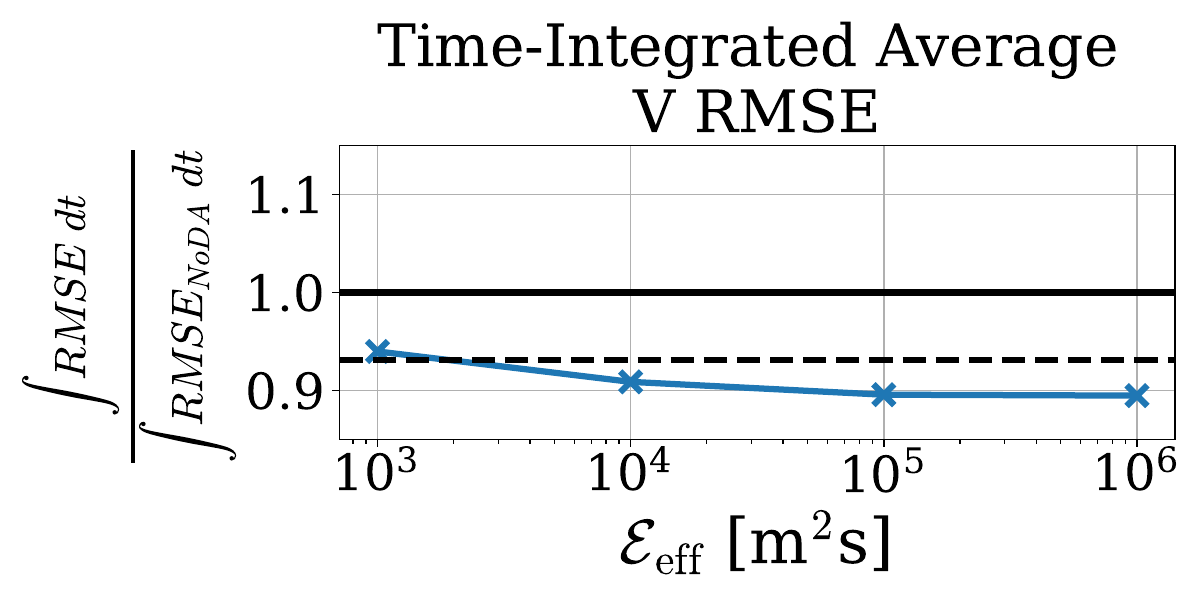}
    \includegraphics[width=0.4\textwidth]{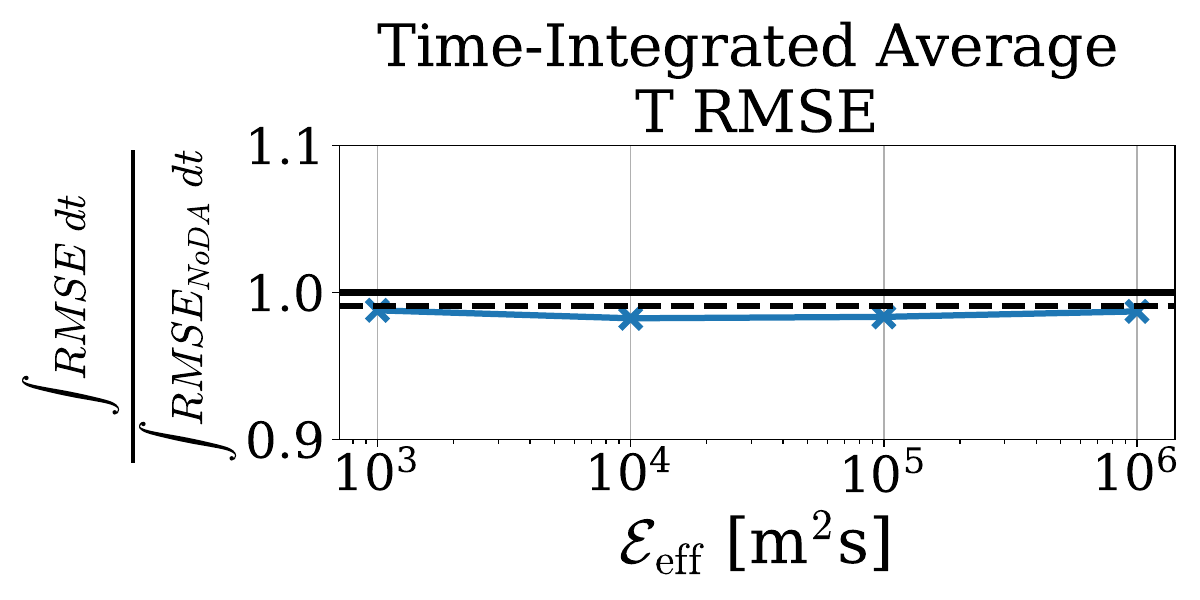}
    \includegraphics[width=0.4\textwidth]{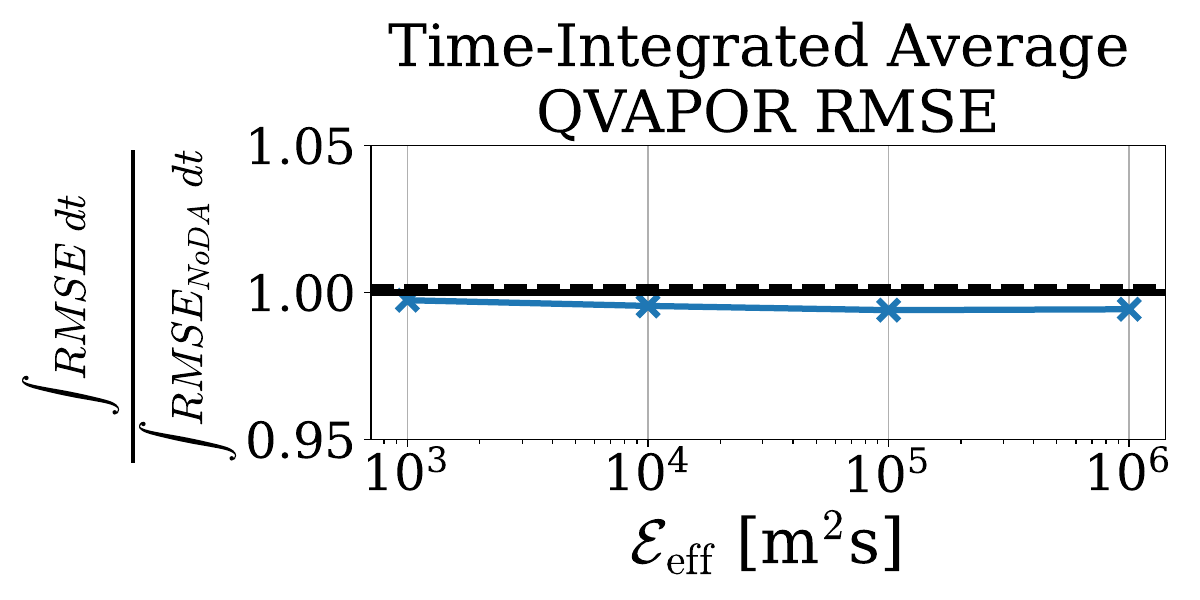}
    \caption{The temporal integral of the curves shown in Figure \ref{fig:rmse_psfc_plot} and \ref{fig:rmse_uvtq_plot}, as a function of muon detector exposure. Values have been normalized to the corresponding temporal integral of RMSE for free model evolution. For comparison, the integrated average RMSE values for no data assimilation (solid black line) and assimilation of a surface pressure point measurement (black dashed line) are also shown. Values lower on the y-axis correspond to an improved ability to predict surface pressure values over a 24-hour period.}
    \label{fig:rmse_tint_exposure}
\end{figure}

Figure \ref{fig:rmse_tint_exposure} shows the curves in figure \ref{fig:rmse_psfc_plot} integrated over the 24-hour assimilation period, and plotted as a function of muon detector exposure. These values have been normalized to NoDA to provide an estimate of the fractional improvement the different assimilated observations offer. As expected, muon detectors with larger exposures produce larger improvements in surface pressure forecasting over the domain considered, up to a limit. Muon data assimilation gives similar performance to PSFC assimilation even for the smallest muon detectors considered ($\mathcal{E}_{\text{eff}} = 10^3$ m$^2$ s), with larger detectors providing over 10\% additional improvement.  

U and V forecasts show similar improvements with muon detector exposure, and in almost all cases are better than assimilation of a surface pressure point measurement. As discussed in the previous sections, improvements in T and QVAPOR forecasts are small, but notably still present for large enough muon detectors. 

While real muon detectors have complicated angular and energy dependence, if we consider an ideal muon detector with flat energy and directional response, we can approximate the effective exposure to be simply:
\begin{equation}
    \mathcal{E}_{\text{eff}} \approx \Delta T \times \overline{A}_{\text{eff}}
    \label{eq:eeff_approx}
\end{equation}
where $\overline{A}_{\text{eff}}$ is the muon detector effective area, averaged over energy and muon arrival direction, and $\Delta T$ is the duration over which the atmospheric muon flux is measured. As $\overline{A}_{\text{eff}}$ will be roughly the physical size of the muon detector, we can use equation~\ref{eq:eeff_approx} to arrive at an estimate for the combination of muon detector size and livetime needed to produce improvements in weather forecasting. If a muon flux observation duration of $\Delta T = 1$ hour is chosen, muon-powered weather forecasts of PSFC could surpass assimilation of PSFC point measurements with a detector as small as 0.27 m$^2$. Alternatively, if the detector size is known, the above constraint can be used to determine the required livetime of muon data: A 1 m$^2$ muon detector would require 16.7 minutes to accumulate enough data to improve weather forecasts more than a surface pressure point measurement, and a 10 m$^2$ muon detector could do the same with only 1.7 minutes of data. 

This detector exposure requirement is remarkably small, and many cosmic ray detectors exceeding the sizes listed above by many orders of magnitude already exist~\cite{Auger, Teshima:1997sc, GUPTA2005311, ayres:in2p3-00704734}. Smaller muon detectors in the 1 to 10 m$^2$ range can also be easily constructed~\cite{kauer2019scintillatorupgradeicetopperformance}, suggesting a potential low-cost avenue for novel weather instrumentation. In either case, meteorological data assimilation of muon flux rates is well within current technological capabilities.  

Muon flux assimilation could also be performed as observations of opportunity using data from existing astroparticle detectors, some of which are even conveniently located for meteorological observations. The IceCube Neutrino Observatory~\cite{Aartsen_2017} is located at the South Pole and could aid in improving atmospheric characterization over Antarctica. P-ONE~\cite{Agostini_2020} is planned to be located off the western coast of Canada, providing an excellent opportunity for improved characterization of the North American jet stream. Both of these detectors are (or will be) a cubic kilometer or larger in volume, and can easily clear the exposure requirement calculated above. 

Importantly, in this study, we have only made use of the total all-sky and all-energy integrated muon flux information. While (as noted in the next section) the atmospheric muon flux does contain directional information, a muon detector does not need to be able to reconstruct individual muon event directions or energies to contribute to improved forecasts, merely the total count over a given time period. This vastly simplifies the muon detector design requirements. Scintillator-based muon detectors of this type can be easily constructed and scaled to the desired size, and in fact this is often done as a component of larger experiments~\cite{kauer2019scintillatorupgradeicetopperformance}, or even as student lab activities~\cite{Axani_2018}.

\subsection{Directionality}
While the studies in this paper primarily focus on assimilating the integrated, all-sky atmospheric muon flux, it should be noted that muon flux measurements contain directional information about the atmospheric density field~\cite{typhoons,PhysRevD.111.023018}. The all-sky integrated muon flux considered in this paper is partially sensitive to directional changes, if those directional changes change the total muon flux. A muon flux excess from one direction will increase the total muon flux as long as there is not a corresponding muon flux deficit in a different direction. This effect likely plays some role in the improvements seen when assimilating muon flux in comparison to a surface pressure point measurement, however this approach is almost certainly not optimal. 

\begin{figure}
    \includegraphics[width=0.45\textwidth]{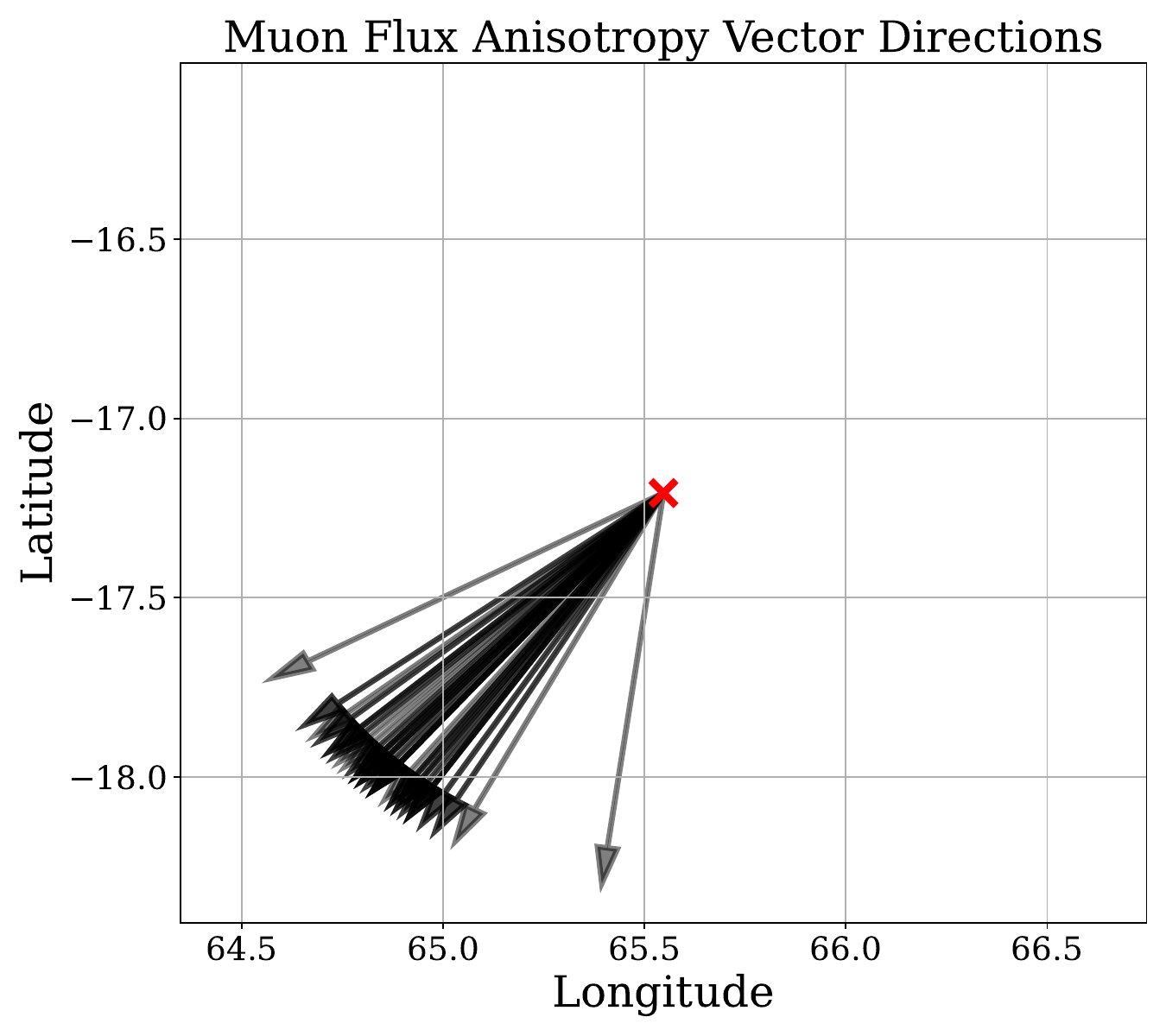}
    \caption{Directions of greatest muon flux anisotropy, for 50 different ensemble members at $t=1$ hour. The direction of the plotted arrows correspond to the direction of greatest muon excess, i.e. the direction of $\phi$ that maximizes $\Phi(\phi)$.}
    \label{fig:avectors}
\end{figure}

Improvement in performance could likely be obtained through explicit assimilation of muon flux from different directions, or some other description of the muon flux anisotropy, provided a muon detector is able to detect changes in the directional profile of the incoming muon flux. Figure~\ref{fig:avectors} shows the direction of the vector of greatest muon flux anisotropy for each ensemble member at $t=1$ hour, showing how different ensemble members produce different muon flux anisotropies. Assimilation of this information is certainly possible, though the procedure for best implementing this in practice is somewhat unclear, and this would further complicate the design of muon detectors attempting to do this. As such, we choose to report our findings using the all-sky integrated muon flux and leave further exploration of the directional muon flux to later work.

\subsection{Effects of Solar Activity}
Solar modulation is known to induce $\approx$1\% level variations on the atmospheric muon flux~\cite{solarmuon}. For comparison, variations between ensemble member simulated muon fluxes are $< 1$\%. If solar modulations of the cosmic ray flux are uncorrelated with local weather, this represents a potentially major systematic that would remove or at least significantly reduce the benefits of muon flux assimilation. 

Notably, solar modulation effects are expected to be periodic, and previous studies of the correlation between cosmic ray rates and solar activity suggests that independent measurements of the solar state could be used to account for the corresponding effect on the atmospheric muon flux~\cite{MAGHRABI20212941}. Similarly, transient extra-atmospheric events such as Forbush decreases could be identified via correlations between distributed arrays of muon detectors at similar geomagnetic latitudes, or via cross-referencing with solar activity and neutron monitor data~\cite{2021MNRAS.503.5675O} allowing for re-calibration of relative muon rates if the disturbance is well-characterized, or masking of the data for the duration of the disturbance otherwise. As the recovery period of Forbush decreases can last hours to days~\cite{KILIFARSKA2020101}, the latter is not ideal, but still results in non-negligible uptime during non-Forbush decrease time periods. 

It should also be noted that previous studies have identified a potential link between strong Forbush decreases and rainfall~\cite{Stozhkov:1995uz,KNIVETON20041135}. If true, this suggests potential utility in the assimilation of uncorrected atmospheric muon flux measurements for long-term weather prediction. While interesting, further exploration of this connection would require a detailed understanding of the processes connecting cosmic rays and global precipitation. This does not appear to be readily available at the time of writing this paper. As such, we leave this investigation for future work.

\section{Conclusion}
The correlation between atmospheric density and atmospheric muon flux is a well-understood phenomenon that has been observed by a variety of particle detectors over the years~\cite{muthunderstorms, muthunderstorms2}, however previous studies of this phenomenon have struggled to find a practical application. In this work, we have shown that atmospheric muon flux measurements can be used to improve meteorological forecasts: assimilation of atmospheric muon flux information leads to improved forecasts of surface pressure, wind velocity, temperature, and humidity. Crucially, these forecast improvements appear to be larger than improvements associated with assimilation of an individual barometric point measurement. 

On the other hand, traditional barometric surface observations are abundant and assimilated with a much richer observational framework. While quantification of the non-locality of muon flux measurements may be required to fully understand the mechanism by which muon detectors can be used to improve weather forecasts, these results seem to indicate that muon flux measurements could be an untapped data source with properties at least on par with existing instruments in the context of atmospheric data assimilation. Further studies will likely be required to test the performance of muon flux assimilation in the context of existing barometric observation networks. 

Basic muon detectors are cheap to build by particle detector standards, but are certainly more expensive than traditional barometers. However, it should be noted that the forecast improvements presented in this work can be achieved with a relatively small muon detector (less than 10 m$^2$, potentially even less than 1 m$^2$), a requirement that is met by a multitude of existing particle detectors around the world. This suggests significant potential for observations of opportunity using the copious amounts of atmospheric muon data collected by detectors such as the IceCube Neutrino Observatory, Pierre Auger Observatory, Telescope Array, GRAPES, and many others.

\section{Acknowledgments}
The authors would like to thank Helen Kershaw, Jeffrey Anderson, and John Beacom for their help and useful discussions throughout the course of this project, as well as Peter Taylor for constructive feedback on plot presentation.

\section*{Appendix A: More about the EnKF}

The ensemble Kalman filter (EnKF) is among the most popular DA methods to assimilate measurements into NWP pipelines ~\cite{Evensen1994a,Burgers1998, Houtekamer1998a, Bishop2001AdaptiveAspects, Pham2001StochasticSystems, Whitaker2002, Anderson2003AFiltering}. Fig \ref{fig:ens_DA_illustration} illustrates the typical workflow of an EnKF-based DA system. The EnKF iteratively incorporate measurements into NWP over time while accounting for both measurement uncertainties and forecast uncertainties. The forecast uncertainties are estimated from probabilistic forecasts, which are obtained by running an ensemble of NWP simulations (forecast ensemble). The EnKF adjusts unmeasured NWP model quantities based on the assimilated measurement using linear regressions linking the measured quantity to the unmeasured quantity. Those relationships are estimated automatically from the forecast ensemble. 

Note that the EnKF is frequently used to assimilate observations that are nonlinearly related to the model state, which includes radar reflectivity and satellite radiances \cite{dowell_etal2011_enkf_reflectivity, Otkin2010, Zhang2016, Chan_etal2022_TMeCSR, MallickJones2022_WoFS_IRDA,  Kugler2023PotentialPrediction}. Such assimilation is made possible through formulating the EnKF in a joint observation-model state vector form \cite{Anderson2003AFiltering}. See Chapters 2 and 3 of \cite{Chan_2022_PhD_Thesis} for an extensive discussion of the joint observation-model state vector formulation of the EnKF.


\section*{Appendix B: Details Regarding the Setup of the WRF Model and the Weather Simulation Ensemble}

Our WRF model setup has 573$\times$267 9-km-wide grid boxes in the horizontal, 45 terrain-following model eta levels ~\cite{SimmonsBurridge_1981_Vertical_Coordinate_systems}, a model top pressure of 2,000 Pa, a 30-second time-step, and utilizes the Lambert map projection. This setup is chosen over one with higher horizontal resolution to limit computational costs. 

Microphysical cloud processes are parameterized using the Thompson scheme ~\cite{Thompson2008ExplicitParameterization}, surface layer processes are parameterized via Mesoscale Model 5 scheme ~\cite{Jimenez2012AFormulation}, land surface processes are parameterized by the Noah land surface model ~\cite{Tewari_etal2004_NoahLand}, boundary layer processes are parameterized using the Yonsei University scheme ~\cite{Hong2006}, longwave radiation processes are represented via the Rapid Radiative Transfer Model for Global Circulation Models ~\cite{Iacono2008}, and shortwave radiation processes are handled using the Goddard scheme \cite{Chou1999}. To reduce computational cost, the radiative forcing are recalculated every 12 time steps. As a 9-km grid spacing is small enough to explicitly resolve mesoscale convective systems ~\cite{Wang2015RegionalResolution}, no cumulus parameterization scheme is used here. In exchange for computational affordability, our setup's 9-km grid spacing is too coarse to resolve individual deep convective updrafts.

All WRF simulations in this study employ initial and boundary conditions (ICBCs) based on a popular atmospheric reanalysis dataset -- the European Center for Medium-range Weather Forecast's Reanalysis version 5 (ERA5; ~\cite{Hersbach2020TheReanalysis}). The conversion of ERA5 data into ICBCs is mediated by the WRF Preprocessing System (WPS) version 4.5.1. The boundary conditions are based on three-hourly data from the ERA5. We used three-hourly data instead of one-hourly data as the ERA5 ensemble used in our ensemble simulations is only available every 3 hours. All simulations are initialized on 18 February 2023 at 1200 Universal Time Coordinate (UTC), and the WRF model is allowed to ``spin-up'' for 12-hours before performing any experiments. In other words, all of our experiments begin on 19 February 2023 at 0000 UTC.

The NR is generated by running WRF using ICBCs derived from the ERA5 control member (also known as the "reanalysis"). This use of the control member ensures that the NR's simulated TC Freddy is reasonably similar to the actual TC Freddy. 

For DA, we use 50-member ensembles of WRF simulations (i.e., the second kind of weather simulations). These 50 WRF simulations are constructed from the ERA5's 10 members. To produce the 50 ICBCs needed to run the 50 WRF simulations, a three-step procedure is used. First, we obtain 10 ensemble state perturbations  $\left\lbrace \boldsymbol{x'_1},\, \boldsymbol{x'_2},\,\dots,\,\boldsymbol{x'_{10}} \right\rbrace$ from the 10 ERA5 members $\left\lbrace \boldsymbol{x_1},\, \boldsymbol{x_2},\,\dots,\,\boldsymbol{x_{10}} \right\rbrace$ via
\begin{equation}
    \boldsymbol{x'_n} = \boldsymbol{x_n} - \dfrac{1}{10}\sum^{10}_{m=1} \boldsymbol{x_m}
    \,\,\,\,\,\,
    \forall
    \,\, n=1,2,\dots,10.
\end{equation}
Efficient scalable covariance-conserving resampling is then applied on $\left\lbrace \boldsymbol{x'_1},\, \boldsymbol{x'_2},\,\dots,\,\boldsymbol{x'_{10}} \right\rbrace$ to construct 41 additional ensemble perturbations $\left\lbrace \boldsymbol{x'_{11}},\, \boldsymbol{x'_{12}},\,\dots,\,\boldsymbol{x'_{51}} \right\rbrace$ \cite{Chan2020BGEnKF, Chan_etal2022_BGEnKF_WRF_OSSE, Chan2024_PESEGC}. In the third and final step, the necessary 50 ICBCs are obtained by combining $\left\lbrace \boldsymbol{x'_1},\, \boldsymbol{x'_2},\,\dots,\,\boldsymbol{x'_{51}} \right\rbrace$ with the ERA5 control member $\boldsymbol{x_{\text{ERA5}}}$ via
\begin{equation}\label{eq:icbc_offset}
    \boldsymbol{x^r_n} = \boldsymbol{x_{\text{ERA5}}} + \boldsymbol{x'_n} + \boldsymbol{x'_{51}} 
    \,\,\,\,\,\,
    \forall
    \,\, n=1,2,\dots,50.
\end{equation}
Note that offset term $\boldsymbol{x'_{51}}$ in Eq. \eqref{eq:icbc_offset} is added to ensure that average of the 50 ICBCs is different from the ICBCs used by the nature run. 

\section*{Appendix C: Details of DART Setup}

The WRF model variables updated by our DART setup are: three-dimensional wind velocities, geopotential heights, surface pressures, water vapor mass mixing ratios, cloud water mass mixing ratios, rain water mass mixing ratios, snow mass mixing ratios, cloud ice mass mixing ratios, graupel mass mixing ratios, skin temperatures, and potential temperatures. To help maintain appropriate spread-to-error ratios in the WRF ensembles, 80\% relaxation to prior spread (RTPS; \cite{Whitaker2008}) is applied on the analysis ensemble. 

To suppress the deleterious impacts of sampling errors on the DA performance, spatial localization using the Gaspari-Cohn fifth order rational function is applied onto the Kalman gain \cite{Gaspari1999}. Based on examining the spatial correlations between atmospheric muon measurements and surface pressure, a horizontal radius of influence of 640 km is chosen for horizontal localization. In other words, DART adjusts model variables within a 640 km radius around a measurement.  A vertical radius of influence of 0.75 scale height is chosen for vertical localization. Future work can further optimize the radii of influence.

\section*{Appendix D: Extended Discussion of OSSE Results}

\begin{figure}
    \centering
    \includegraphics[width=0.5\textwidth]{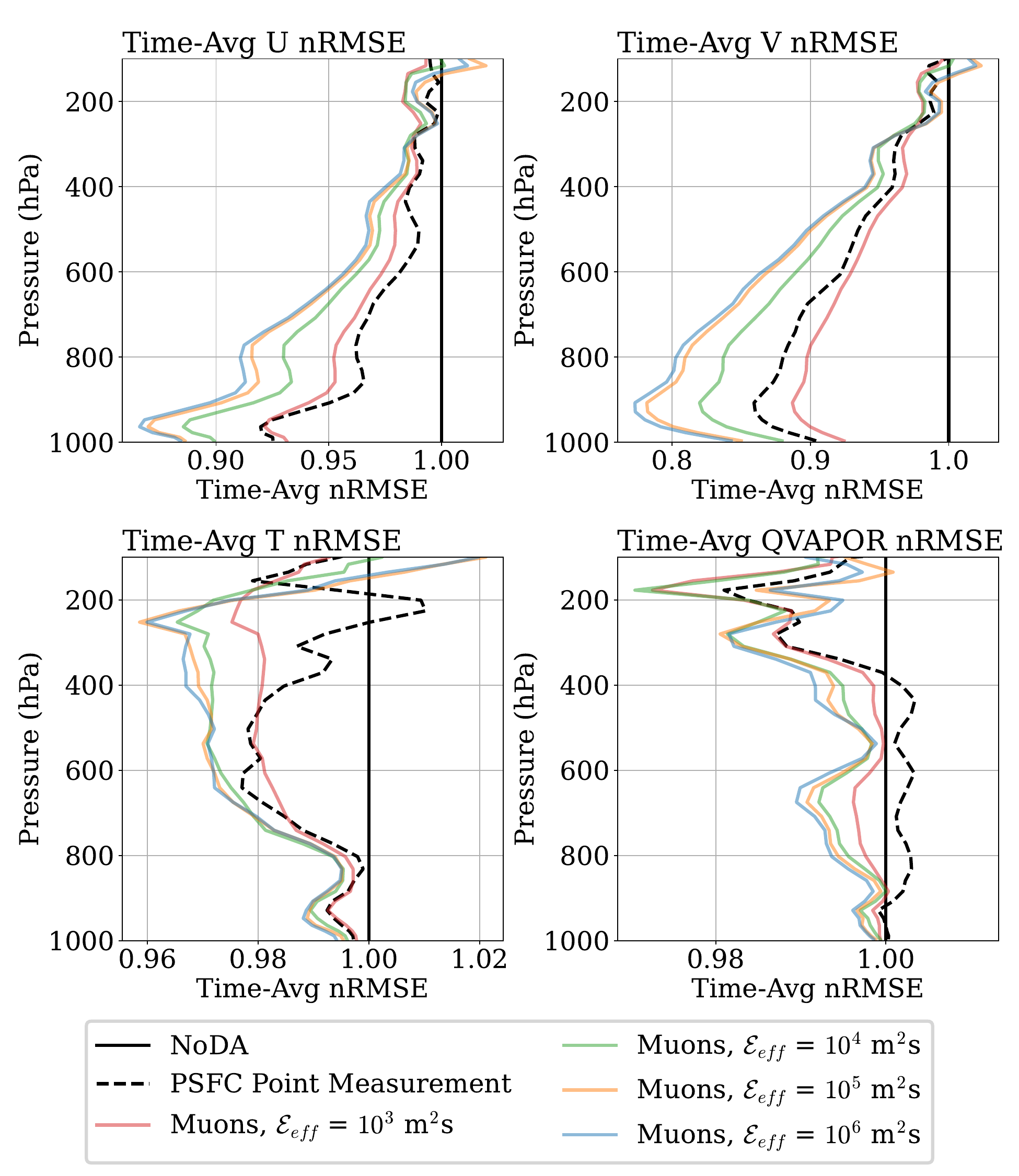}
    \caption{Plots of time-averaged nRMSEs as functions of $p$ for U (top left), V (top right), T (bottom left), QVAPOR (bottom right). }
    \label{fig:rmse_uvtq_plvl_plot}
\end{figure}

We also dissected the impacts of assimilating muon fluxes and PSFC measurements by examining time-averaged nRMSEs as a function of $p$ (Figure \ref{fig:rmse_uvtq_plvl_plot}). Consistent with the earlier discussion surrounding Figure \ref{fig:rmse_uvtq_plot}, Figure \ref{fig:rmse_uvtq_plvl_plot} indicates that the assimilation of muon fluxes generally improved the U, V, T and QVAPOR fields and those improvements can exceed those resulting from assimilating PSFC. The new information revealed by Figure \ref{fig:rmse_uvtq_plvl_plot} is that muon flux DA consistently improved U, V, T and QVAPOR across most tropospheric model levels (200--1000 hPa). Furthermore, the advantages of assimilating muon flux measurements with exposures exceeding $10^3$ m$^2$s over PSFC measurements also occurs consistently for U, V, T and QVAPOR within most tropospheric model levels. The assimilation of muon flux measurements with exposures of $10^3$ m$^2$s also resulted in better time-averaged QVAPOR nRMSEs over most tropospheric model levels than the assimilation of PSFC measurements. As such, Figure \ref{fig:rmse_uvtq_plvl_plot} further substantiates the potential for muon flux measurements to improve NWP and the fact that these measurements may have more powerful impacts on NWP than PSFC measurements.

\clearpage
\bibliography{apssamp, references_joseph}

@article{LECHMANN2021103842,
title = {Muon tomography in geoscientific research – A guide to best practice},
journal = {Earth-Science Reviews},
volume = {222},
pages = {103842},
year = {2021},
issn = {0012-8252},
doi = {https://doi.org/10.1016/j.earscirev.2021.103842},
url = {https://www.sciencedirect.com/science/article/pii/S0012825221003433},
author = {Alessandro Lechmann and David Mair and Akitaka Ariga and Tomoko Ariga and Antonio Ereditato and Ryuichi Nishiyama and Ciro Pistillo and Paola Scampoli and Fritz Schlunegger and Mykhailo Vladymyrov}
}

@misc{tilav2019seasonal,
      title={Seasonal variation of atmospheric muons in IceCube}, 
      author={Serap Tilav and Thomas K. Gaisser and Dennis Soldin and Paolo Desiati},
      year={2019},
      eprint={1909.01406},
      archivePrefix={arXiv},
      primaryClass={astro-ph.HE}
}

@article{typhoons,
author = {Tanaka, Hiroyuki and Gluyas, Jon and Holma, Marko and Joutsenvaara, J. and Kuusiniemi, Pasi and Leone, Giovanni and Lo Presti, D. and Matsushima, Jun and Oláh, László and Steigerwald, Sara and Thompson, Lee and Usoskin, Ilya and Poluianov, Stepan and Varga, Dezső and Yokota, Yusuke},
year = {2022},
month = {10},
pages = {16710},
title = {Atmospheric muography for imaging and monitoring tropic cyclones},
volume = {12},
journal = {Scientific Reports},
doi = {10.1038/s41598-022-20039-4}
}

@article{PhysRevD.111.023018,
  title = {Effect of tornadic supercell thunderstorms on the atmospheric muon flux},
  author = {Luszczak, William and Orf, Leigh},
  journal = {Phys. Rev. D},
  volume = {111},
  issue = {2},
  pages = {023018},
  numpages = {12},
  year = {2025},
  month = {Jan},
  publisher = {American Physical Society},
  doi = {10.1103/PhysRevD.111.023018},
  url = {https://link.aps.org/doi/10.1103/PhysRevD.111.023018}
}

@ARTICLE{h3acitation,
       author = {{Gaisser}, Thomas K.},
        title = "{Spectrum of cosmic-ray nucleons, kaon production, and the atmospheric muon charge ratio}",
      journal = {Astroparticle Physics},
         year = 2012,
        month = jul,
       volume = {35},
       number = {12},
        pages = {801-806},
          doi = {10.1016/j.astropartphys.2012.02.010},
archivePrefix = {arXiv},
       eprint = {1111.6675},
 primaryClass = {astro-ph.HE},
       adsurl = {https://ui.adsabs.harvard.edu/abs/2012APh....35..801G}
}

@article{sibyll,
    author = "Riehn, Felix and Dembinski, Hans P. and Engel, Ralph and Fedynitch, Anatoli and Gaisser, Thomas K. and Stanev, Todor",
    title = "{The hadronic interaction model SIBYLL 2.3c and Feynman scaling}",
    eprint = "1709.07227",
    archivePrefix = "arXiv",
    primaryClass = "hep-ph",
    doi = "10.22323/1.301.0301",
    journal = "PoS",
    volume = "ICRC2017",
    pages = "301",
    year = "2018"
}

@article{Auger,
title = "{The Pierre Auger Cosmic Ray Observatory}",
journal = {Nuclear Instruments and Methods in Physics Research Section A: Accelerators, Spectrometers, Detectors and Associated Equipment},
volume = {798},
pages = {172-213},
year = {2015},
issn = {0168-9002},
doi = {https://doi.org/10.1016/j.nima.2015.06.058},
url = {https://www.sciencedirect.com/science/article/pii/S0168900215008086},
}

@inproceedings{Teshima:1997sc,
    author = "Teshima, M.",
    collaboration = "Telescope Array",
    title = "{Telescope Array project}",
    booktitle = "{32nd Rencontres de Moriond: High-Energy Phenomena in Astrophysics}",
    pages = "217--222",
    year = "1997"
}

@article{GUPTA2005311,
title = {GRAPES-3—A high-density air shower array for studies on the structure in the cosmic-ray energy spectrum near the knee},
journal = {Nuclear Instruments and Methods in Physics Research Section A: Accelerators, Spectrometers, Detectors and Associated Equipment},
volume = {540},
number = {2},
pages = {311-323},
year = {2005},
issn = {0168-9002},
doi = {https://doi.org/10.1016/j.nima.2004.11.025},
url = {https://www.sciencedirect.com/science/article/pii/S016890020402426X},
author = {S.K. Gupta and Y. Aikawa and N.V. Gopalakrishnan and Y. Hayashi and N. Ikeda and N. Ito and A. Jain and A.V. John and S. Karthikeyan and S. Kawakami and T. Matsuyama and D.K. Mohanty and P.K. Mohanty and S.D. Morris and T. Nonaka and A. Oshima and B.S. Rao and K.C. Ravindran and M. Sasano and K. Sivaprasad and B.V. Sreekantan and H. Tanaka and S.C. Tonwar and K. Viswanathan and T. Yoshikoshi}
}

@techreport{ayres:in2p3-00704734,
  TITLE = {{The NOvA Technical Design Report}},
  AUTHOR = {Ayres, D.S. and others},
  URL = {https://in2p3.hal.science/in2p3-00704734},
  PAGES = {600},
  YEAR = {2007},
  MONTH = Oct,
  HAL_ID = {in2p3-00704734},
  HAL_VERSION = {v1},
}

@article{Aartsen_2017,
   title={The IceCube Neutrino Observatory: instrumentation and online systems},
   volume={12},
   ISSN={1748-0221},
   url={http://dx.doi.org/10.1088/1748-0221/12/03/P03012},
   DOI={10.1088/1748-0221/12/03/p03012},
   number={03},
   journal={Journal of Instrumentation},
   publisher={IOP Publishing},
   author={Aartsen, M.G. and others},
   year={2017},
   month=mar, pages={P03012–P03012} }

@article{Agostini_2020,
   title={The Pacific Ocean Neutrino Experiment},
   volume={4},
   ISSN={2397-3366},
   url={http://dx.doi.org/10.1038/s41550-020-1182-4},
   DOI={10.1038/s41550-020-1182-4},
   number={10},
   journal={Nature Astronomy},
   publisher={Springer Science and Business Media LLC},
   author={Agostini, Matteo and Böhmer, Michael and Bosma, Jeff and Clark, Kenneth and Danninger, Matthias and Fruck, Christian and Gernhäuser, Roman and Gärtner, Andreas and Grant, Darren and Henningsen, Felix and Holzapfel, Kilian and Huber, Matthias and Jenkyns, Reyna and Krauss, Carsten B. and Krings, Kai and Kopper, Claudio and Leismüller, Klaus and Leys, Sally and Macoun, Paul and Meighen-Berger, Stephan and Michel, Jan and Moore, Roger and Morley, Mike and Padovani, Paolo and Papp, Laszlo and Pirenne, Benoit and Qiu, Chuantao and Rea, Immacolata Carmen and Resconi, Elisa and Round, Adrian and Ruskey, Albert and Spannfellner, Christian and Traxler, Michael and Turcati, Andrea and Yanez, Juan Pablo},
   year={2020},
   month=sep, pages={913–915} }

@article{Jourde_2016,
   title={Monitoring temporal opacity fluctuations of large structures with muon radiography: a calibration experiment using a water tower},
   volume={6},
   ISSN={2045-2322},
   url={http://dx.doi.org/10.1038/srep23054},
   DOI={10.1038/srep23054},
   number={1},
   journal={Scientific Reports},
   publisher={Springer Science and Business Media LLC},
   author={Jourde, Kevin and Gibert, Dominique and Marteau, Jacques and de Bremond d’Ars, Jean and Gardien, Serge and Girerd, Claude and Ianigro, Jean-Christophe},
   year={2016},
   month=mar }

@article{muthunderstorms,
  title = {Variations of muon flux in the atmosphere during thunderstorms},
  author = {Karapetyan, G. G.},
  journal = {Phys. Rev. D},
  volume = {89},
  issue = {9},
  pages = {093005},
  numpages = {8},
  year = {2014},
  month = {May},
  publisher = {American Physical Society},
  doi = {10.1103/PhysRevD.89.093005},
  url = {https://link.aps.org/doi/10.1103/PhysRevD.89.093005}
}

@article{muthunderstorms2,
title = {Studies of Thunderstorm Events Based on the Data of Muon Hodoscope URAGAN and Meteorological Radar DMRL-C},
journal = {Physics Procedia},
volume = {74},
pages = {486-492},
year = {2015},
note = {Fundamental Research in Particle Physics and Cosmophysics},
issn = {1875-3892},
doi = {https://doi.org/10.1016/j.phpro.2015.09.239},
url = {https://www.sciencedirect.com/science/article/pii/S1875389215014388},
author = {A.V. Kozyrev and N.S. Barbashina and T.A. Belyakova and J.B. Pavlyukov and A.A. Petrukhin and N.I. Serebryannik and V.V. Shutenko and I.I. Yashin}
}

@misc{kauer2019scintillatorupgradeicetopperformance,
      title={The Scintillator Upgrade of IceTop: Performance of the prototype array}, 
      author={Matt Kauer and Thomas Huber and Delia Tosi and Chris Wendt},
      year={2019},
      eprint={1908.09860},
      archivePrefix={arXiv},
      primaryClass={astro-ph.HE},
      url={https://arxiv.org/abs/1908.09860}, 
}

@article{Axani_2018,
   title={The CosmicWatch Desktop Muon Detector: a self-contained, pocket sized particle detector},
   volume={13},
   ISSN={1748-0221},
   url={http://dx.doi.org/10.1088/1748-0221/13/03/P03019},
   DOI={10.1088/1748-0221/13/03/p03019},
   number={03},
   journal={Journal of Instrumentation},
   publisher={IOP Publishing},
   author={Axani, S.N. and Frankiewicz, K. and Conrad, J.M.},
   year={2018},
   month=mar, pages={P03019–P03019} }

@article{Alameddine_2020,
doi = {10.1088/1742-6596/1690/1/012021},
url = {https://dx.doi.org/10.1088/1742-6596/1690/1/012021},
year = {2020},
month = {dec},
publisher = {IOP Publishing},
volume = {1690},
number = {1},
pages = {012021},
author = {Alameddine, J-M and Soedingrekso, J and Sandrock, A and Sackel, M and Rhode, W},
title = {PROPOSAL: A library to propagate leptons and high energy photons},
journal = {Journal of Physics: Conference Series}
}

@Article{instruments6040078,
AUTHOR = {Borja, Cristian and Ávila, Carlos and Roque, Gerardo and Sánchez, Manuel},
TITLE = {Atmospheric Muon Flux Measurement near Earth’s Equatorial Line},
JOURNAL = {Instruments},
VOLUME = {6},
YEAR = {2022},
NUMBER = {4},
ARTICLE-NUMBER = {78},
URL = {https://www.mdpi.com/2410-390X/6/4/78},
ISSN = {2410-390X},
DOI = {10.3390/instruments6040078}
}

@article{solarmuon,
author = {Mubashir, A. and Ashok, A. and Bourgeois, A. G. and Chien, Y. T. and Connors, M. and Potdevin, E. and He, X. and Martens, P. and Mikler, A. and Perera, A. G. U. and Sadykov, V. and Sarsour, M. and Sharma, D. and Tiwari, C.},
title = {Muon Flux Variations Measured by Low-Cost Portable Cosmic Ray Detectors and Their Correlation With Space Weather Activity},
journal = {Journal of Geophysical Research: Space Physics},
volume = {128},
number = {12},
pages = {e2023JA031943},
doi = {https://doi.org/10.1029/2023JA031943},
url = {https://agupubs.onlinelibrary.wiley.com/doi/abs/10.1029/2023JA031943},
eprint = {https://agupubs.onlinelibrary.wiley.com/doi/pdf/10.1029/2023JA031943},
note = {e2023JA031943 2023JA031943},
year = {2023}
}

@article{MAGHRABI20212941,
title = {Correlation analyses between solar activity parameters and cosmic ray muons between 2002 and 2012 at high cutoff rigidity},
journal = {Advances in Space Research},
volume = {68},
number = {7},
pages = {2941-2952},
year = {2021},
issn = {0273-1177},
doi = {https://doi.org/10.1016/j.asr.2021.05.016},
url = {https://www.sciencedirect.com/science/article/pii/S0273117721004361},
author = {A. Maghrabi and A. Aldosari and M. Almutairi}
}

@ARTICLE{2021MNRAS.503.5675O,
       author = {{Okike}, O. and {Alhassan}, J.~A. and {Iyida}, E.~U. and {Chukwude}, A.~E.},
        title = "{A comparison of catalogues of Forbush decreases identified from individual and a network of neutron monitors: a critical perspective}",
      journal = {mnras},
         year = {2021},
        month = {jun},
       volume = {503},
       number = {4},
        pages = {5675-5691},
          doi = {10.1093/mnras/stab680},
       adsurl = {https://ui.adsabs.harvard.edu/abs/2021MNRAS.503.5675O}
}

@incollection{KILIFARSKA2020101,
title = {Chapter 5 - Galactic cosmic rays and solar particles in Earth's atmosphere},
editor = {Natalya A. Kilifarska and Volodymyr G. Bakhmutov and Galyna V. Melnyk},
booktitle = {The Hidden Link between Earth's Magnetic Field and Climate},
publisher = {Elsevier},
pages = {101-131},
year = {2020},
isbn = {978-0-12-819346-4},
doi = {https://doi.org/10.1016/B978-0-12-819346-4.00005-X},
url = {https://www.sciencedirect.com/science/article/pii/B978012819346400005X},
author = {Natalya A. Kilifarska and Volodymyr G. Bakhmutov and Galyna V. Melnyk}
}

@article{Stozhkov:1995uz,
    author = "Stozhkov, Yu. I. and Bazilevskaya, G. A. and Makhmutov, V. S. and Svirzhevsky, N. S. and Zullo, J. and Martin, I. M. and Pellegrino, G. Q. and Pinto, H. S. and Bezerra, P. C. and Turtelli, A.",
    title = "{Rainfalls during great Forbush decreases}",
    doi = "10.1007/BF02508564",
    journal = "Nuovo Cim. C",
    volume = "18",
    pages = "335--341",
    year = "1995"
}

@article{KNIVETON20041135,
title = {Precipitation, cloud cover and Forbush decreases in galactic cosmic rays},
journal = {Journal of Atmospheric and Solar-Terrestrial Physics},
volume = {66},
number = {13},
pages = {1135-1142},
year = {2004},
note = {SPECIAL - Space Processes and Electrical Changes in Atmospheric L ayers},
issn = {1364-6826},
doi = {https://doi.org/10.1016/j.jastp.2004.05.010},
url = {https://www.sciencedirect.com/science/article/pii/S1364682604001087},
author = {D.R Kniveton}
}

@article{Chan_etal2022_TMeCSR,
    title = {{A High‐Resolution Tropical Mesoscale Convective System Reanalysis (TMeCSR)}},
    year = {2022},
    journal = {Journal of Advances in Modeling Earth Systems},
    author = {Chan, Man-Yau and Chen, Xingchao and Leung, L. Ruby},
    number = {9},
    month = {9},
    volume = {14},
    publisher = {American Geophysical Union (AGU)},
    doi = {10.1029/2021ms002948},
    issn = {1942-2466}
}

@article{Hong2006,
    title = {{A new vertical diffusion package with an explicit treatment of entrainment processes}},
    year = {2006},
    journal = {Monthly Weather Review},
    author = {Hong, Song You and Noh, Yign and Dudhia, Jimy},
    doi = {10.1175/MWR3199.1},
    issn = {00270644}
}

@article{Chou1999,
    title = {{A Solar Radiation Parameterization Atmospheric Studies}},
    year = {1999},
    journal = {Technical Report Series on Global Modeling and Data Assimilation},
    author = {Chou, Ming-Dah and Suarez, Max J}
}

@article{Chan2020BGEnKF,
    title = {{An efficient bi-Gaussian ensemble Kalman filter for satellite infrared radiance data assimilation}},
    year = {2020},
    journal = {Monthly Weather Review},
    author = {Chan, Man-Yau and Anderson, Jeffrey L. and Chen, Xingchao},
    doi = {10.1175/mwr-d-20-0142.1},
    issn = {0027-0644}
}

@article{SimmonsBurridge_1981_Vertical_Coordinate_systems,
    title = {{An energy and angular-momentum conserving vertical finite- difference scheme and hybrid vertical coordinates.}},
    year = {1981},
    journal = {Monthly Weather Review},
    author = {Simmons, A. J. and Burridge, D. M.},
    number = {4},
    volume = {109},
    doi = {10.1175/1520-0493(1981)109<0758:AEAAMC>2.0.CO;2},
    issn = {00270644}
}

@article{Burgers1998,
    title = {{Analysis scheme in the ensemble Kalman filter}},
    year = {1998},
    journal = {Monthly Weather Review},
    author = {Burgers, Gerrit and Jan van Leeuwen, Peter and Evensen, Geir and Van Leeuwen, Peter Jan and Evensen, Geir},
    number = {6},
    month = {6},
    pages = {1719--1724},
    volume = {126},
    publisher = {American Meteorological Society},
    url = {http://journals.ametsoc.org/doi/10.1175/1520-0493(1998)126%3C1719:ASITEK%3E2.0.CO;2},
    doi = {10.1175/1520-0493(1998)126<1719:ASITEK>2.0.CO;2},
    issn = {00270644}
}

@book{Kalnay2003,
    title = {{Atmospheric Modelling, Data Assimilation}},
    year = {2003},
    booktitle = {Quarterly Journal of the Royal Meteorological Society},
    author = {Kalnay, E},
    number = {592},
    pages = {2441--2442},
    volume = {129},
    publisher = {Cambridge University Press},
    url = {http://doi.wiley.com/10.1256/00359000360683511},
    isbn = {9780521791793},
    doi = {10.1256/00359000360683511},
    issn = {1477870X},
    pmid = {25246403},
    arxivId = {arXiv:1011.1669v3}
}

@misc{BOM_2023_TC_Summary,
    title = {{Australia’s 2022–23 northern wet season}},
    year = {2023},
    author = {{Bureau of Meteorology}},
    month = {6},
    url = {https://www.bom.gov.au/climate/current/season/tropics/archive/202304.summary.shtml}
}

@article{Otkin2010,
    title = {{Clear and cloudy sky infrared brightness temperature assimilation using an ensemble Kalman filter}},
    year = {2010},
    journal = {Journal of Geophysical Research Atmospheres},
    author = {Otkin, Jason A.},
    doi = {10.1029/2009JD013759},
    issn = {01480227}
}

@article{Morrison_etal2020_Review_Cloud_Microphysics,
    title = {{Confronting the Challenge of Modeling Cloud and Precipitation Microphysics}},
    year = {2020},
    journal = {Journal of Advances in Modeling Earth Systems},
    author = {Morrison, Hugh and van Lier-Walqui, Marcus and Fridlind, Ann M. and Grabowski, Wojciech W. and Harrington, Jerry Y. and Hoose, Corinna and Korolev, Alexei and Kumjian, Matthew R. and Milbrandt, Jason A. and Pawlowska, Hanna and Posselt, Derek J. and Prat, Olivier P. and Reimel, Karly J. and Shima, Shin Ichiro and van Diedenhoven, Bastiaan and Xue, Lulin},
    number = {8},
    volume = {12},
    doi = {10.1029/2019MS001689},
    issn = {19422466}
}

@article{Gaspari1999,
    title = {{Construction of correlation functions in two and three dimensions}},
    year = {1999},
    journal = {Quarterly Journal of the Royal Meteorological Society},
    author = {Gaspari, Gregory and Cohn, Stephen E.},
    number = {554},
    month = {1},
    pages = {723--757},
    volume = {125},
    publisher = {Wiley},
    url = {http://doi.wiley.com/10.1002/qj.49712555417},
    doi = {10.1256/smsqj.55416},
    issn = {00359009}
}

@book{Evensen_etal2022_DA_textbook,
    title = {{Data Assimilation Fundamentals}},
    year = {2022},
    booktitle = {Springer Textbooks in Earth Sciences, Geography and Environment},
    author = {Evensen, Geir and Vossepoel, Femke Cathelijne and van Leeuwen, Peter Jan}
}

@article{Houtekamer1998a,
    title = {{Data assimilation using an ensemble Kalman filter technique}},
    year = {1998},
    journal = {Monthly Weather Review},
    author = {Houtekamer, P. L. and Mitchell, Herschel L.},
    number = {3},
    pages = {796--811},
    volume = {126},
    doi = {10.1175/1520-0493(1998)126<0796:DAUAEK>2.0.CO;2},
    issn = {00270644}
}

@article{Whitaker2008,
    title = {{Ensemble data assimilation with the NCEP global forecast system}},
    year = {2008},
    journal = {Monthly Weather Review},
    author = {Whitaker, Jeffrey S. and Hamill, Thomas M. and Wei, Xue and Song, Yucheng and Toth, Zoltan},
    number = {2},
    volume = {136},
    doi = {10.1175/2007MWR2018.1},
    issn = {00270644}
}

@article{Whitaker2002,
    title = {{Ensemble data assimilation without perturbed observations}},
    year = {2002},
    journal = {Monthly Weather Review},
    author = {Whitaker, Jeffrey S. and Hamill, Thomas M.},
    number = {7},
    month = {7},
    pages = {1913--1924},
    volume = {130},
    publisher = {American Meteorological Society},
    url = {http://journals.ametsoc.org/doi/10.1175/1520-0493(2002)130%3C1913:EDAWPO%3E2.0.CO;2},
    doi = {10.1175/1520-0493(2002)130<1913:EDAWPO>2.0.CO;2},
    issn = {00270644}
}

@article{dowell_etal2011_enkf_reflectivity,
    title = {{Ensemble Kalman Filter Assimilation of Radar Observations of the 8 May 2003 Oklahoma City Supercell: Influences of Reflectivity Observations on Storm-scale Analyses}},
    year = {2011},
    journal = {Monthly Weather Review},
    author = {Dowell, David C. and Wicker, Louis J. and Snyder, Chris},
    number = {1},
    month = {1},
    pages = {272--294},
    volume = {139},
    url = {http://journals.ametsoc.org/doi/10.1175/2010MWR3438.1},
    doi = {10.1175/2010MWR3438.1},
    issn = {00270644}
}

@incollection{ECMWF_2024_IFS_DA,
    title = {{IFS Documentation CY49R1 - Part II: Data Assimilation}},
    year = {2024},
    booktitle = {IFS Documentation CY49R1},
    author = {{ECMWF}},
    chapter = {2},
    month = {2},
    publisher = {ECMWF},
    url = {&nbsp;},
    doi = {10.21957/105cb1333c}
}

@article{MallickJones2022_WoFS_IRDA,
    title = {{Impact of adaptively thinned GOES-16 all-sky radiances in an ensemble Kalman filter based WoFS}},
    year = {2022},
    journal = {Atmospheric Research},
    author = {Mallick, Swapan and Jones, Thomas A},
    pages = {106304},
    volume = {277},
    url = {https://www.sciencedirect.com/science/article/pii/S0169809522002903},
    doi = {https://doi.org/10.1016/j.atmosres.2022.106304},
    issn = {0169-8095}
}

@article{Anderson2003AFiltering,
    title = {{A Local Least Squares Framework for Ensemble Filtering}},
    year = {2003},
    journal = {Monthly Weather Review},
    author = {Anderson, Jeffrey L.},
    number = {4},
    month = {4},
    pages = {634--642},
    volume = {131},
    url = {http://journals.ametsoc.org/doi/10.1175/1520-0493(2003)131<0634:ALLSFF>2.0.CO;2},
    doi = {10.1175/1520-0493(2003)131<0634:ALLSFF>2.0.CO;2},
    issn = {0027-0644}
}

@article{Jimenez2012AFormulation,
    title = {{A revised scheme for the WRF surface layer formulation}},
    year = {2012},
    journal = {Monthly Weather Review},
    author = {Jim{\'{e}}nez, Pedro A. and Dudhia, Jimy and Gonz{\'{a}}lez-Rouco, J. Fidel and Navarro, Jorge and Mont{\'{a}}vez, Juan P. and Garc{\'{i}}a-Bustamante, Elena},
    number = {3},
    volume = {140},
    doi = {10.1175/MWR-D-11-00056.1},
    issn = {00270644}
}

@article{Bishop2001AdaptiveAspects,
    title = {{Adaptive Sampling with the Ensemble Transform Kalman Filter. Part I: Theoretical Aspects}},
    year = {2001},
    journal = {Monthly Weather Review},
    author = {Bishop, Craig H. and Etherton, Brian J. and Majumdar, Sharanya J.},
    number = {3},
    month = {3},
    pages = {420--436},
    volume = {129},
    publisher = {American Meteorological Society},
    url = {http://journals.ametsoc.org/doi/10.1175/1520-0493(2001)129%3C0420:ASWTET%3E2.0.CO;2 https://journals.ametsoc.org/view/journals/mwre/129/3/1520-0493_2001_129_0420_aswtet_2.0.co_2.xml},
    doi = {10.1175/1520-0493(2001)129<0420:ASWTET>2.0.CO;2}
}

@article{Anderson2001AnAssimilation,
    title = {{An ensemble adjustment Kalman filter for data assimilation}},
    year = {2001},
    journal = {Monthly Weather Review},
    author = {Anderson, Jeffrey L.},
    number = {12},
    month = {12},
    pages = {2884--2903},
    volume = {129},
    publisher = {American Meteorological Society},
    url = {http://journals.ametsoc.org/doi/10.1175/1520-0493(2001)129%3C2884:AEAKFF%3E2.0.CO;2},
    doi = {10.1175/1520-0493(2001)129<2884:AEAKFF>2.0.CO;2},
    issn = {00270644}
}

@incollection{Fletcher2017ApplicationsGeosciences,
    title = {{Applications of Data Assimilation in the Geosciences}},
    year = {2017},
    booktitle = {Data Assimilation for the Geosciences},
    author = {Fletcher, Steven J.},
    pages = {887--916},
    publisher = {Elsevier},
    doi = {10.1016/b978-0-12-804444-5.00023-4}
}

@article{Thompson2008ExplicitParameterization,
    title = {{Explicit forecasts of winter precipitation using an improved bulk microphysics scheme. Part II: Implementation of a new snow parameterization}},
    year = {2008},
    journal = {Monthly Weather Review},
    author = {Thompson, Gregory and Field, Paul R. and Rasmussen, Roy M. and Hall, William D.},
    number = {12},
    month = {12},
    pages = {5095--5115},
    volume = {136},
    url = {http://journals.ametsoc.org/doi/10.1175/2008MWR2387.1},
    doi = {10.1175/2008MWR2387.1},
    issn = {00270644}
}

@inproceedings{Tewari_etal2004_NoahLand,
    title = {{Implementation and verification of the unified noah land surface model in the WRF model}},
    year = {2004},
    booktitle = {Bulletin of the American Meteorological Society},
    author = {Tewari, M. and Chen, F. and Wang, W. and Dudhia, J. and LeMone, M. A. and Mitchell, K. and Ek, M. and Gayno, G. and Wegiel, J. and Cuenca, R. H.},
    issn = {00030007}
}

@article{Chan2024_PESEGC,
    title = {{Improving Ensemble Data Assimilation through Probit-space Ensemble Size Expansion for Gaussian Copulas (PESE-GC)}},
    year = {2024},
    journal = {Nonlinear Processes in Geophysics},
    author = {Chan, Man-Yau},
    url = {https://npg.copernicus.org/articles/31/287/2024/npg-31-287-2024.pdf},
    doi = {10.5194/npg-31-287-2024}
}

@phdthesis{Chan_2022_PhD_Thesis,
    title = {{Improving the analysis and forecast of tropical mesoscale convective systems through advancing the ensemble data assimilation of geostationary satellite infrared radiance observations}},
    year = {2022},
    author = {Chan, Man-Yau},
    month = {10},
    url = {https://etda.libraries.psu.edu/catalog/21389mxc98},
    school = {The Pennsylvania State University},
    address = {State College}
}

@article{BAMS2024_TC_Freddy,
    title = {{New WMO Certified Tropical Cyclone Duration Extreme TC Freddy (04 February to 14 March 2023) Lasting for 36.0 Days}},
    year = {2024},
    journal = {Bulletin of the American Meteorological Society},
    author = {Earl-Spurr, Craig and Langlade, Sébastien and Krahenbuhl, Daniel and Aberson, Sim D. and Brunet, Manola and Chan, Johnny and Fogarty, Chris and Landsea, Christopher W. and Trewin, Blair and Velden, Christopher and Balling, Robert C. and Cerveny, Randall S.},
    number = {12},
    volume = {105},
    doi = {10.1175/BAMS-D-24-0071.1},
    issn = {15200477}
}

@article{Zhang2016,
    title = {{Potential impacts of assimilating all-sky infrared satellite radiances from GOES-R on convection-permitting analysis and prediction of tropical cyclones}},
    year = {2016},
    journal = {Geophysical Research Letters},
    author = {Zhang, Fuqing and Minamide, Masashi and Clothiaux, Eugene E.},
    number = {6},
    pages = {2954--2963},
    volume = {43},
    doi = {10.1002/2016GL068468},
    issn = {19448007}
}

@article{Iacono2008,
    title = {{Radiative forcing by long-lived greenhouse gases: Calculations with the AER radiative transfer models}},
    year = {2008},
    journal = {Journal of Geophysical Research Atmospheres},
    author = {Iacono, Michael J. and Delamere, Jennifer S. and Mlawer, Eli J. and Shephard, Mark W. and Clough, Shepard A. and Collins, William D.},
    doi = {10.1029/2008JD009944},
    issn = {01480227}
}

@misc{temperature,
    title = {{Sea Surface Temperature Trends As A Function Of Latitude Bands By Roger A. Pielke Sr. and Bob Tisdale | Climate Science: Roger Pielke Sr.}},
    year = {2012},
    author = {{Rpielke}},
    url = {https://pielkeclimatesci.wordpress.com/2012/07/11/sea-surface-temperature-trends-as-a-function-of-latitude-bands-by-roger-a-pielke-sr-and-bob-tisdale/}
}

@article{Evensen1994a,
    title = {{Sequential data assimilation with a nonlinear quasi-geostrophic model using Monte Carlo methods to forecast error statistics}},
    year = {1994},
    journal = {Journal of Geophysical Research},
    author = {Evensen, Geir},
    number = {C5},
    pages = {10143--10162},
    volume = {99},
    url = {https://doi.org/10.1029/94JC00572 http://doi.wiley.com/10.1029/94JC00572},
    doi = {10.1029/94JC00572},
    issn = {01480227}
}

@article{Hersbach2020TheReanalysis,
    title = {{The ERA5 global reanalysis}},
    year = {2020},
    journal = {Quarterly Journal of the Royal Meteorological Society},
    author = {Hersbach, Hans and Bell, Bill and Berrisford, Paul and Hirahara, Shoji and Hor{\'{a}}nyi, András and Mu{\~{n}}oz-Sabater, Joaquín and Nicolas, Julien and Peubey, Carole and Radu, Raluca and Schepers, Dinand and Simmons, Adrian and Soci, Cornel and Abdalla, Saleh and Abellan, Xavier and Balsamo, Gianpaolo and Bechtold, Peter and Biavati, Gionata and Bidlot, Jean and Bonavita, Massimo and De Chiara, Giovanna and Dahlgren, Per and Dee, Dick and Diamantakis, Michail and Dragani, Rossana and Flemming, Johannes and Forbes, Richard and Fuentes, Manuel and Geer, Alan and Haimberger, Leo and Healy, Sean and Hogan, Robin J. and H{\'{o}}lm, Elías and Janiskov{\'{a}}, Marta and Keeley, Sarah and Laloyaux, Patrick and Lopez, Philippe and Lupu, Cristina and Radnoti, Gabor and de Rosnay, Patricia and Rozum, Iryna and Vamborg, Freja and Villaume, Sebastien and Th{\'{e}}paut, Jean Noël},
    doi = {10.1002/qj.3803},
    issn = {1477870X}
}

@article{Chan_etal2022_BGEnKF_WRF_OSSE,
    title = {{The potential benefits of handling mixture statistics via a bi-Gaussian EnKF: tests with all-sky satellite infrared radiances}},
    year = {2023},
    journal = {Journal of Advances in Modeling Earth Systems},
    author = {Chan, Man-Yau and Chen, Xingchao and Anderson, Jeffrey L},
    pages = {},
    url = {https://agupubs.onlinelibrary.wiley.com/doi/epdf/10.1029/2022MS003357?src=getftr},
    doi = {10.1029/2022MS003357}
}

@misc{wmo2024_freddy,
    title = {{Tropical Cyclone Freddy is the longest tropical cyclone on record at 36 days}},
    year = {2024},
    booktitle = {WMO Press Release},
    author = {{World Meteorological Organization}},
    month = {7}
}

@article{Kugler2023PotentialPrediction,
    title = {{Potential impact of all-sky assimilation of visible and infrared satellite observations compared with radar reflectivity for convective-scale numerical weather prediction}},
    year = {2023},
    journal = {Quarterly Journal of the Royal Meteorological Society},
    author = {Kugler, Lukas and Anderson, Jeffrey L. and Weissmann, Martin},
    number = {757},
    volume = {149},
    doi = {10.1002/qj.4577},
    issn = {1477870X}
}

@article{Wang2015RegionalResolution,
    title = {{Regional simulation of the october and november MJO events observed during the CINDY/DYNAMO field campaign at gray zone resolution}},
    year = {2015},
    journal = {Journal of Climate},
    author = {Wang, Shuguang and Sobel, Adam H. and Zhang, Fuqing and Qiang Sun, Y. and Yue, Ying and Zhou, Lei},
    number = {6},
    pages = {2097--2119},
    volume = {28},
    isbn = {10.1175/JCLI-D-14-00294.1},
    doi = {10.1175/JCLI-D-14-00294.1},
    issn = {08948755}
}

@article{Pham2001StochasticSystems,
    title = {{Stochastic Methods for Sequential Data Assimilation in Strongly Nonlinear Systems}},
    year = {2001},
    journal = {Monthly Weather Review},
    author = {Pham, Dinh Tuan},
    number = {5},
    month = {5},
    pages = {1194--1207},
    volume = {129},
    publisher = {American Meteorological Society},
    url = {http://journals.ametsoc.org/doi/10.1175/1520-0493(2001)129<1194:SMFSDA>2.0.CO;2},
    doi = {10.1175/1520-0493(2001)129<1194:SMFSDA>2.0.CO;2},
    issn = {0027-0644}
}

@article{Anderson2009TheFacility,
    title = {{The data assimilation research testbed a community facility}},
    year = {2009},
    journal = {Bulletin of the American Meteorological Society},
    author = {Anderson, Jeffrey L. and Hoar, Tim and Raeder, Kevin and Liu, Hui and Collins, Nancy and Torn, Ryan and Avellano, Avelino},
    number = {9},
    month = {9},
    pages = {1283--1296},
    volume = {90},
    url = {https://journals.ametsoc.org/doi/10.1175/2009BAMS2618.1},
    doi = {10.1175/2009BAMS2618.1},
    issn = {00030007}
}

@article{E,
    title = {{ 0 B U  {\^{a}} {\textregistered}¦ H {${}^\circ$}  h u E}}
}

\end{document}